\documentclass[sn-mathphys]{sn-jnl}

\jyear{2023}%

\usepackage{amsmath,verbatim}	
\usepackage{amssymb}	
\usepackage{amsfonts}
\usepackage{amsthm}				
\usepackage{mathtools} 

\usepackage{xcolor}	

\renewcommand{\vec}[1]{\boldsymbol{#1}}
\let\oldhat\hat
\renewcommand{\hat}[1]{\oldhat{\boldsymbol{#1}}}

\newcommand{\de}{\ensuremath{\partial}}						
\newcommand{\dee}{\ensuremath{\textrm{d}}}
\newcommand{\pd}[2]{\ensuremath{ \frac{\de #1}{\de #2}}}
\newcommand{\pdf}[1]{\ensuremath{ \frac{\de}{\de #1}}}

\newcommand{\fd}[2]{\ensuremath{ \frac{\dee #1}{\dee #2}}}

\newcommand{\inty}[4]{\ensuremath{ \int_{#1}^{#2} \! #3 \, \dee#4 }}

\newcommand{\ip}[2]{\ensuremath{ \left< \left. #1 \right\vert #2 \right> } }

\newcommand{\db}{{\rm d}}

\newcommand{\aw}[1]{{{\color{blue}{AW: #1}}}}

\newtheorem{assumption}{Assumption}
\newtheorem{remark}{Remark}
\newtheorem{theorem}{Theorem}

\newtheorem{lemma}{Lemma}
\newtheorem{proposition}{Proposition}

\numberwithin{lemma}{section}
\numberwithin{example}{section}
\numberwithin{equation}{section}
\numberwithin{theorem}{section}
\numberwithin{corollary}{section}
\numberwithin{remark}{section}
\numberwithin{definition}{section}
\numberwithin{assumption}{section}
\numberwithin{figure}{section}
\newtheorem*{definition*}{Definition}		
\newtheorem*{assumption*}{Assumption}
\newtheorem*{remark*}{Remark}
\newtheorem*{theorem*}{Theorem}
\newtheorem*{lemma*}{Lemma}
\newtheorem*{proposition*}{Proposition}
\newtheorem*{corollary*}{Corollary}
\newtheorem*{example*}{Example}
\newtheorem*{conjecture*}{Conjecture}

\let\Re\relax
\DeclareMathOperator{\Re}{Re}
\let\Im\relax
\DeclareMathOperator{\Im}{Im}
\let\Tr\relax
\DeclareMathOperator{\Tr}{Tr}

\mathtoolsset{showonlyrefs}

\title[Kubo formula with dissipation]{Mathematical aspects of the Kubo formula for electrical conductivity with dissipation}

\author[1]{\fnm{Alexander B.} \sur{Watson}}\email{watso860@umn.edu}

\author[2]{\fnm{Dionisios} \sur{Margetis}}\email{diom@umd.edu}

\author*[1]{\fnm{Mitchell} \sur{Luskin}}\email{luskin@umn.edu}

\affil*[1]{\orgdiv{Mathematics Department}, \orgname{University of Minnesota Twin Cities}, \orgaddress{\street{206 Church St SE}, \city{Minneapolis}, \postcode{55455}, \state{MN}, \country{United States}}}

\affil[2]{\orgdiv{Department of Mathematics, and Institute for Physical Science \& Technology}, \orgname{University of Maryland}, \orgaddress{\city{College Park}, \postcode{20742}, \state{MD}, \country{United States}}}

\abstract{In this expository article, we present a systematic formal derivation of the Kubo formula for the 
linear-response current due to a time-harmonic electric field applied to non-interacting, spinless charged particles in a finite volume in the quantum setting. We model dissipation in a transparent way by assuming a sequence of scattering events occurring at random-time intervals modeled by a Poisson distribution. By taking the large-volume limit, we derive special cases of the formula for free electrons, continuum and tight-binding periodic systems, and the nearest-neighbor tight-binding model of graphene. We present the analogous formalism with dissipation to derive the Drude conductivity of classical free particles.
}

\begin{document}

\maketitle

\tableofcontents

\section{Introduction}
\subsection{Motivation} The partial differential equations modeling solid mechanics, fluid mechanics, and electromagnetics include material moduli (heat, magnetic, and electric susceptibility, transport coefficients, etc.) that describe the macroscopic response to a small perturbation.  A general approach to formulating the linear response of a physical system at equilibrium was proposed by Ryobo Kubo in the seminal papers \cite{Ryogo1957,Ryogo19572}.  


In this article, we give a systematic formal derivation of the Kubo formula for the electrical conductivity of non-interacting, spinless charged particles, e.g., spinless\footnote{Although electrons have spin, many phenomena of practical interest can be captured by treating electrons as spinless and then taking into account spin degeneracy at the end of the calculation.} electrons in a material. This Kubo formula plays a key role in explanations of many important phenomena such as the quantum Hall effect \cite{1982ThoulessKohmotoNightingaledenNijs,Ryogo1957,Ryogo19572,Allen,fradkin_2013,Tong,Tong2016,ashcroft_mermin,kaxiras_joannopoulos_2019,dresselhaus,Schulz-Baldes1998,Lein,Bouclet2005,Klein_annals,Teufel2020,Henheik2021,Cances2017a,Bru2017,MargetisWatsonLuskin2023}.\footnote{For further discussion of these references, see Section \ref{sec:related}.}

The specific form of Kubo's formula we derive is general enough to apply to a wide variety of models of interest. However, we also show, in detail, how this form reduces to well-known simplified forms in common special cases. Our goal is to stimulate studies of this subject in the context of modern materials applications. We believe that this research direction can inspire exciting questions in mathematical modeling and numerical analysis.


\subsection{General Kubo formula} We start by presenting the Kubo formula we derive without details, to communicate the main ideas. Consider a system of negatively-charged, non-interacting, spinless quantum particles in $d$ dimensions, that are initially at equilibrium. Applying a time-harmonic electric field $\vec{E}(\omega) e^{- i \omega t}$ to this system will excite an electrical current density at the same temporal frequency $\vec{J}(\omega) e^{- i \omega t}$. Formally, we can expand the current density in powers of the field as
\begin{equation} \label{eq:naive_linear_response}
    \vec{J}(\omega) = \delta_{\omega 0} \vec{J}_{\text{eq}} + \sigma(\omega) \vec{E}(\omega) + O(\vec{E}^2),
\end{equation}
where $\omega$ denotes the temporal frequency of the field, and $\vec{J}_{\text{eq}}$ denotes the current density in equilibrium. The Kubo formula is an expression for the linear coefficient in this expansion, $\sigma$, known as the electrical conductivity. 


To state the formula, consider, more specifically, $N$ non-interacting spinless particles confined to a subset $\Omega \subset \mathbb{R}^d$ with volume $\vert\Omega\vert$. In the quantum setting, this system can be described by the density matrix $\rho$, acting on a single-particle Hilbert space $\mathcal{H}$, evolving according to the von Neumann equation with Hamiltonian $H$. We assume that, at equilibrium, $\rho$ equals the Fermi-Dirac distribution function $\Phi(H)$. Then, the Kubo formula can be formulated  as \cite{Schulz-Baldes1998}
\begin{equation} \label{eq:standard_Kubo}
    \sigma_{l m}(\omega) = - \frac{e^2}{\hbar^2} \tilde{\Tr}\left\{ (\de_l H) \left( \mathcal{L}_H - i \omega + \Gamma \right)^{-1} \de_m \Phi(H) \right\}, \quad 1 \leq l, m \leq d~.
\end{equation}
In the above, $-e, \hbar, \Gamma$ denote the  electron charge ($e>0$), reduced Planck's constant, and inverse mean scattering time, respectively; $\tilde{\text{Tr}}$ denotes the trace density in $\mathcal{H}$, defined by the trace divided by the volume $\vert \Omega \vert$; $\de_l := - i [X_l,\cdot]$ denotes derivation\footnote{Here we use the mathematical notion of derivation as an operation which generalizes the ordinary derivative. Derivations are linear operations which satisfy Leibniz's law; see, e.g., \cite{bourbaki_algebra}.} with respect to the components, $X_l,$ where $1 \leq l \leq d$, of the position operator $\vec{X}$; and $\mathcal{L}_H := \frac{i}{\hbar}[H,\cdot]$ denotes the Liouvillian of $H$, where $[A,B] := AB - BA$ is the commutator of the operators $A, B$. The notation $\de_l$ for the derivation is natural, since $- i [X_l,\cdot]$ acts as $\pdf{k_l}$ in the Fourier and Bloch domains where $\hbar k_l$ is the $l$-th component of the momentum vector.

It is important to emphasize the generality of formula \eqref{eq:standard_Kubo}. It applies equally well to continuum Schr\"odinger Hamiltonians as to tight-binding Hamiltonians, and does not require periodicity. However, we will show that it contains many standard forms of the Kubo formula as special cases. For example, the Drude conductivity for free particles \eqref{eq:free_Kubo}, and the Kubo formula for periodic systems including both Drude and ``interband'' contributions \eqref{eq:periodic_Kubo_0}.


\subsection{Formalism}  \label{sec:formalism}

The first systematic derivation of a linear relation between electric field and current in metals was introduced by Drude \cite{Drude1,Drude2}. Drude's model gave qualitative agreement with some experimental results, but also gave incorrect predictions for the specific heat~\cite{ashcroft_mermin}, for example, since it considers electrons to be classical particles. The emergence of quantum mechanics led to early improvements to the Drude model, most simply by replacing the Boltzmann-Maxwell statistics by Fermi-Dirac statistics \cite{sommerfeld28}. More systematic quantum theories that utilize the band structure of crystals were proposed by Bloch \cite{bloch28} and others have continued to be actively developed in the physics literature.


To the best of our knowledge, the Kubo formula in the form \eqref{eq:standard_Kubo}, with $\omega = 0$, was first introduced and derived by Bellissard, van Elst, and Schulz-Baldes. The formula played a key role in their investigation of the quantum Hall effect \cite{Bellissard1994}, because it allowed for a formulation of the Hall conductivity even in the presence of magnetic fields and random disorder breaking translational symmetry. In particular, they were able to prove that, under appropriate conditions, the large volume limit $\vert \Omega \vert \rightarrow \infty$ of \eqref{eq:standard_Kubo} converges to $\frac{e^2}{2 \pi \hbar}$ times an integer. The formula \eqref{eq:standard_Kubo} with $\omega \neq 0$ was then introduced and rigorously justified in \cite{Schulz-Baldes1998}. The works \cite{Bellissard1994} and \cite{Schulz-Baldes1998} use the $C^*$-algebra formalism, which is well-suited for studying quantum systems without translational symmetry. Other discussions of Kubo's formula across the physics and mathematics literature include, for example, \cite{Allen,fradkin_2013,Tong,Tong2016,ashcroft_mermin,kaxiras_joannopoulos_2019,dresselhaus,Lein,Bouclet2005,Klein_annals,Teufel2020,Henheik2021,ElgartSchlein,Bachmann_quantization,Bachmann_extendedspins}. We discuss these works in more detail in Section \ref{sec:related}.

The first goal of this work is to systematically derive \eqref{eq:naive_linear_response}-\eqref{eq:standard_Kubo}, following the main ideas of \cite{Bellissard1994}-\cite{Schulz-Baldes1998}, but avoiding the $C^*$-algebra formalism, and without striving for mathematical rigor. We hope that this presentation will make the core ideas of these works accessible to a wider array of practitioners across mathematics, physics, and engineering.

The main ideas of this calculation can be summarized as follows. The current density is initially defined as the trace of $\vec{j} \rho$, where $\vec{j} := \vec{\nabla} H$ is the current density observable, where $\vec{\nabla} := ( \de_l )_{1 \leq l \leq d}$. In the absence of dissipation, $\rho$ evolves according to the von Neumann equation, a differential equation with time-periodic coefficients, which models the effect of the applied field $\vec{E}$ together with the unperturbed system Hamiltonian $H$. Dissipation is, then, modeled by a sequence of scattering events which return $\rho$ to its equilibrium distribution. These events occur at random time intervals, whose lengths are modeled by a Poisson distribution with rate $\Gamma$. With these ingredients in place, the current density $\vec{J}$ appearing in \eqref{eq:naive_linear_response} is defined as an average as the number of scattering events tends to infinity (in an appropriate sense); see \eqref{eq:average_constant_case} and \eqref{eq:time_average_timedep}. The fact that scattering and the parameter $\Gamma$ are introduced at the level of the von Neumann equation makes our treatment of dissipation transparent. Equations \eqref{eq:naive_linear_response}-\eqref{eq:standard_Kubo} are calculated by a careful analysis of the propagator of the von Neumann equation for $\rho$ between scattering events, for which we find a simple representation; see \eqref{eq:simple_propagator}. We provide details of our modeling assumptions and results in Section \ref{sec:results}, before giving the detailed derivation in Section \ref{sec:quantum_derivation}.





\subsection{Simplified forms of the Kubo formula} The second goal of the present work is to clarify how \eqref{eq:standard_Kubo} unifies various well-known simplified forms of the Kubo formula, including those for free particles and for particles in a periodic potential. We provide detailed derivations of these forms starting from the trace formula \eqref{eq:standard_Kubo} by taking a large volume limit $\vert \Omega \vert \rightarrow \infty$. Although we have not found these details in the literature, we are likely not the first to carry out such calculations. These calculations are provided in Section \ref{sec:reductions}.

In a similar vein, we also show that the formalism discussed in Section \ref{sec:formalism} for deriving \eqref{eq:naive_linear_response}-\eqref{eq:standard_Kubo} has a natural classical analog. For example, we can replace the density matrix $\rho$ by a classical phase space density, and its evolution under the von Neumann equation by evolution under the classical Liouville equation. We apply this classical formalism to the simplest case, of free particles, in order to derive the Drude conductivity, in Section \ref{sec:classical_formalism}. To the best of our knowledge, this calculation is original to this work. We expect it is possible to derive our classical formalism from the quantum one via a semiclassical limit, but we do not attempt this here.

For free particles, treated either quantum-mechanically via a large-volume limit of \eqref{eq:standard_Kubo}, or classically, we derive \eqref{eq:naive_linear_response}, with $\sigma$ given by the isotropic Drude conductivity \cite{Drude1,Drude2,ashcroft_mermin}
\begin{equation} \label{eq:free_Kubo}
    \sigma (\omega) = \frac{e^2 \tilde{N}}{m ( \Gamma - i \omega )}, \quad \tilde{N} = \frac{N}{\vert\Omega\vert}.
\end{equation}
Here, $\tilde{N}, m$ denote the electron density and mass, respectively. We derive \eqref{eq:free_Kubo} from \eqref{eq:standard_Kubo} in Section \ref{sec:free}.

For particles in a periodic potential, modeled either by a continuum Hamiltonian (Section \ref{sec:periodic}) or tight-binding model (Section \ref{sec:tight-binding}), we derive \eqref{eq:naive_linear_response}, with 
\begin{equation} \label{eq:sigma_two_terms}
    \sigma(\omega) = \sigma^{\text{D}}(\omega) + \sigma^{\text{R}}(\omega).
\end{equation}
The first term in \eqref{eq:sigma_two_terms}, $\sigma^{\text{D}}$, known as the Drude conductivity, generalizes \eqref{eq:free_Kubo} as
\begin{equation} \label{eq:Drude_0}
    \sigma^{\text{D}}_{lm}(\omega) = \frac{ e^2 \tilde{N} ( m^{\text{eff}} )^{-1}_{lm} }{ \Gamma - i \omega }.
\end{equation}
Here, $m^{\text{eff}}$ is a possibly anisotropic effective mass tensor, most conveniently defined through the entries of its inverse as
\begin{align} \label{eq:effective_mass}
        &( m^{\text{eff}} )^{-1}_{lm} := - \frac{1}{\hbar^2 (2 \pi)^d \tilde{N}} \sum_{n \in \mathbb{N}_{> 0}} \inty{\Gamma^*}{}{}{\vec{k}}  \\ \nonumber
        &\qquad \fd{\Phi}{E}(E_n(\vec{k})) \ip{ \chi_{n}(\vec{k}) }{ \left( \pd{H}{k_l}(\vec{k}) \right) \chi_{n}(\vec{k}) } \ip{ \chi_{n}(\vec{k}) }{ \left( \pd{H}{k_m}(\vec{k}) \right) \chi_n(\vec{k}) }.
\end{align}
Here, $1 \leq l, m \leq d$, the symbol $\Gamma^*$ denotes a unit cell of the reciprocal lattice (Brillouin zone), and $H(\vec{k})$, $\chi_n$, and $E_n$ denote the Bloch Hamiltonian, periodic Bloch functions, and Bloch band functions, respectively. 

If every band of $H(\vec{k})$ is simple (separated from all other bands by gaps), formula \eqref{eq:effective_mass} can be written in two more suggestive forms. The first one,
\begin{equation} \label{eq:alternative_form}
    ( m^{\text{eff}} )^{-1}_{lm} = - \frac{1}{\hbar^2 (2 \pi)^d \tilde{N}} \sum_{n \in \mathbb{N}_{> 0}} \inty{\Gamma^*}{}{ \fd{\Phi}{E}(E_n(\vec{k})) \pd{E_n}{k_l}(\vec{k}) \pd{E_n}{k_m}(\vec{k}) }{\vec{k}},
\end{equation}
emphasizes derivatives of the Bloch bands, which equal the group velocity of wave-packets formed by superposing Bloch functions; see, e.g., \cite{panati_spohn_teufel_1,panati_spohn_teufel_2,E2013,Watson2017}. The second one, obtained from \eqref{eq:alternative_form} via integration by parts,
\begin{equation} \label{eq:second_deriv}
    ( m^{\text{eff}} )^{-1}_{lm} = \frac{1}{\hbar^2 (2 \pi)^d \tilde{N}} \sum_{n \in \mathbb{N}_{> 0}} \inty{\Gamma^*}{}{ \Phi(E_n(\vec{k})) \frac{\partial^2 E_n}{\partial k_l \partial k_m}(\vec{k}) }{\vec{k}},
\end{equation}
emphasizes second derivatives of the Bloch bands, which are related to the ``effective mass'' of wave-packets formed by superposing Bloch functions; see, e.g., \cite{1996PoupaudRinghofer,allaire_piatnitski,2006Sparber}. Expression \eqref{eq:second_deriv} reduces straightforwardly to the diagonal matrix $\frac{1}{m}$ when there is just one band, $E(\vec{k}) = \frac{\hbar^2}{2 m} \vert \vec{k} \vert^2$, so that $\tilde{N} = \frac{1}{(2 \pi)^d} \inty{\Gamma^*}{}{ \Phi(E(\vec{k})) }{\vec{k}}$. 

The second term in \eqref{eq:sigma_two_terms}, $\sigma^{\text{R}}$, has no classical analog. It is known as the regular, or interband, conductivity
\begin{align} \label{eq:regular2}
        &\sigma^{\text{R}}_{lm}(\omega) := - \frac{e^2}{\hbar^2 (2 \pi)^d} \sum_{n \neq n' \in \mathbb{N}_{> 0}} \inty{\Gamma^*}{}{}{\vec{k}}  \\ \nonumber
        &\frac{ \left( \Phi(E_{n'}(\vec{k})) - \Phi(E_n(\vec{k})) \right) \ip{ \chi_{n}(\vec{k}) }{ \left( \pd{H}{k_l}(\vec{k}) \right) \chi_{n'}(\vec{k}) } \ip{ \chi_{n'}(\vec{k}) }{ \left( \pd{H}{k_m}(\vec{k}) \right) \chi_n(\vec{k}) } }{(E_{n'}(\vec{k}) - E_n(\vec{k}))\left(\frac{i}{\hbar}\left(E_{n'}(\vec{k}) - E_n(\vec{k})\right) - i \omega + \Gamma\right)}.
    \end{align}
This term is responsible for materials' optical properties; see, e.g., \cite{dresselhaus}. An explicit asymptotic formula for the interband conductivity of the one-dimensional Su-Schrieffer-Heeger model \cite{Su1979}, which demonstrates the importance of interband ``resonances'', was derived in \cite{MargetisWatsonLuskin2023}. 

A convenient expression of the Kubo formula for periodic systems, which takes into account both the Drude and regular contributions, is \cite{ashcroft_mermin,1982ThoulessKohmotoNightingaledenNijs}
\begin{align} \label{eq:periodic_Kubo_0}
        &\sigma_{lm}(\omega) = - \frac{e^2}{\hbar^2 (2 \pi)^d} \sum_{n,n' \in \mathbb{N}_{> 0}} \inty{\Gamma^*}{}{}{\vec{k}}  \\ \nonumber
        &\frac{ \left( \Phi(E_{n'}(\vec{k})) - \Phi(E_{n}(\vec{k})) \right) \ip{ \chi_{n}(\vec{k}) }{ \left( \pd{H}{k_l} (\vec{k}) \right) \chi_{n'}(\vec{k}) } \ip{ \chi_{n'}(\vec{k}) }{ \left( \pd{H}{k_m}(\vec{k}) \right) \chi_n(\vec{k}) } }{(E_{n'}(\vec{k}) - E_{n}(\vec{k}))\left(\frac{i}{\hbar} \left( E_{n'}(\vec{k}) - E_n(\vec{k}) \right) - i \omega + \Gamma\right)}.
\end{align}

Similar reductions of the electrical conductivity are also possible for certain aperiodic systems under ergodicity assumptions; see Section \ref{sec:related}. In the specific case of the nearest-neighbor tight-binding model of graphene, analytical formulas are available for $H(\vec{k})$, $E_n$, and $\chi_n$, so that the Drude and regular conductivities can be expressed through explicit formulas. We review this case in Section \ref{sec:graphene}.

\subsection{Related work} \label{sec:related}

Discussions of Kubo's formula in the physics literature, in addition to Kubo's original papers \cite{Ryogo1957,Ryogo19572}, can be found in \cite{Allen,fradkin_2013,Tong,Tong2016,ashcroft_mermin,kaxiras_joannopoulos_2019,dresselhaus}. Mathematically rigorous approaches to the Kubo formula, other than \cite{Bellissard1994,Schulz-Baldes1998}, include \cite{Lein,Bouclet2005,Klein_annals,Teufel2020,Henheik2021,ElgartSchlein,Bachmann_quantization,Bachmann_extendedspins,Bru2017}. We do not review these works in detail since, other than \cite{Bellissard1994,Schulz-Baldes1998}, they do not model dissipation. Note that some of these works, e.g., \cite{Teufel2020,Henheik2021,Bachmann_quantization,Bachmann_extendedspins,Bru2017}, treat the full many-body Schr\"odinger equation with interactions. 

As discussed in Section \ref{sec:formalism}, large volume limits of \eqref{eq:standard_Kubo} exist even for some aperiodic systems under ergodicity assumptions \cite{Bellissard1994,Schulz-Baldes1998,Bouclet2005,Klein_annals}. We do not address this case in the present work, except to mention that another case where such limits exist can be found in models of incommensurate twisted bilayers; see, e.g., \cite{Cances2017a,Massatt2020,kubocomp20}. 

\subsection{Structure of the paper}

We present our precise modeling assumptions, especially the treatment of dissipation, and our results in the quantum case, in Section \ref{sec:results}. We then derive the general quantum Kubo formula in Section \ref{sec:quantum_derivation}. We derive simplifications of the Kubo formula in the large volume limit for free particles, particles in periodic potentials modeled by continuum and tight-binding models, and the nearest-neighbor tight-binding model of graphene, in Section \ref{sec:reductions}. We derive the classical Drude conductivity, via an analogous formalism, in Section \ref{sec:classical_formalism}. We give a conclusion in Section \ref{sec:conclusion}.



\section{Assumptions and main results} \label{sec:results}

In this section, we introduce our formalism for deriving the general Kubo formula \eqref{eq:naive_linear_response}-\eqref{eq:standard_Kubo}, and state our result precisely. In particular, we introduce the Fermi-Dirac distribution, von Neumann equation for the evolution of the density matrix, and our treatment of dissipation via random scattering events with rate $\Gamma$. Details of the calculations which lead to the Kubo formula \eqref{eq:general_Kubo} will be given in Section \ref{sec:quantum_derivation}. \\


\noindent \emph{Equilibrium density matrix}  We consider the density matrix $\rho$ of a system of non-interacting electrons in a region $\Omega \subset \mathbb{R}^d$, with volume $\vert\Omega\vert$. We denote the effective single-particle Hamiltonian in the absence of applied fields by $H$, and the single-particle Hilbert space by $\mathcal{H}$. The equilibrium density matrix is, then, given by the Fermi-Dirac distribution function 
\begin{equation} \label{eq:Fermi-Dirac}
    \Phi(H) := \left\{ e^{ \beta(H - \mu) } + 1 \right\}^{-1}.
\end{equation}
Here, $\beta := (k_\text{B} T)^{-1}$ is inverse temperature $T$ scaled by Boltzmann's constant $k_\text{B}$, and $\mu$ is the chemical potential. The number of particles in the region $\Omega$, $N$, and the particle density, $\tilde{N}$, are given by 
\begin{equation} 
    N := \Tr \Phi(H), \quad \tilde{N} := \frac{N}{\vert \Omega \vert},
\end{equation}
where $\Tr$ denotes the trace over $\mathcal{H}$.  \\

\noindent \emph{Evolution of density matrix under applied field} When a (possibly time-dependent) electric field $\vec{{E}}$ is applied for $t \ge 0$, and in the absence of dissipation (see below), the density matrix evolves according to the von Neumann equation:
\begin{equation} \label{eq:vonNeumann}
    \fd{\rho}{t} = - \mathcal{L}_{H_{\vec{{E}}}} \rho = - \mathcal{L}_H \rho + \frac{e \vec{{E}}}{\hbar} \cdot \vec{\nabla} \rho, \quad t \geq 0, \quad \rho(0) = \Phi(H).
\end{equation}
Here, $H_{\vec{{E}}}$ denotes the Hamiltonian of the system subject to the applied electric field,  
\begin{equation}
    H_{\vec{{E}}} := H + e \vec{{E}} \cdot \vec{X}.
\end{equation}
In addition, we denote the Liouvillian operators of $H$ and $H_{\vec{{E}}}$ by
\begin{equation}
    \mathcal{L}_H := \frac{i}{\hbar} [ H, \cdot ], \quad \mathcal{L}_{H_{\vec{{E}}}} :=  \frac{i}{\hbar} [ H_{\vec{{E}}}, \cdot ] = \mathcal{L}_H - \frac{e}{\hbar} \vec{{E}} \cdot \vec{\nabla};
\end{equation}
and introduce notation for the derivation
\begin{equation} \label{eq:derivation}
    \vec{\nabla} := - i [ \vec{X}, \cdot ], \quad ( \de_m )_{1 \leq m \leq d} := ( - i [ X_m , \cdot ] )_{1 \leq m \leq d},
\end{equation}
where $\vec{X} = ( X_m )_{1 \leq m \leq d}$ denotes the position operator. \\

\noindent \emph{Modeling of dissipation} We introduce dissipation as follows. We assume that scattering events occur at a random sequence of times $0 = t_0 < t_1 < ...$, such that the differences $\tau_n := t_{n+1} - t_n > 0$ are modeled by a Poisson distribution with rate $\Gamma > 0$, so that
\begin{equation} \label{eq:poisson}
    \mathbb{P}(\tau_n \leq \nu) = \inty{0}{\nu}{ \Gamma e^{- \Gamma \tau_n} }{\tau_n}.
\end{equation}
The average time between scattering events is thus $\Gamma^{-1}$. After each scattering event, we assume that the system returns to equilibrium, before evolving again according to \eqref{eq:vonNeumann} in-between scattering events. Hence, we assume that $\rho$ evolves according to (for $n \in \mathbb{N}_{\geq 0}$)
\begin{equation} \label{eq:vonNeumann_scattering}
    \fd{\rho}{t} = - \mathcal{L}_{H_{\vec{{E}}}} \rho = - \mathcal{L}_H \rho + \frac{e \vec{{E}}}{\hbar} \cdot \vec{\nabla} \rho, \quad t_n \leq t < t_{n+1}, \quad \rho(t_n) = \Phi(H).
\end{equation}

\begin{remark}
    The electron scattering can be modeled more accurately using a scattering kernel. The replacement of such a scattering kernel formalism by the above process involving the (scalar) rate $\Gamma$ is known as the relaxation time approximation \cite{ashcroft_mermin,Schulz-Baldes1998}.
\end{remark}

\noindent \emph{Current operator} We invoke procedures of averaging, both over time and over scattering times modeled by \eqref{eq:poisson}, of the current density expectation $\vec{{J}}$. This expectation is defined in terms of the trace density $\tilde{\Tr} := \frac{1}{\vert \Omega \vert} \Tr$, where $\Tr$ is the trace over $\mathcal{H}$, and the current density operator $\vec{j}$. The related expression is
\begin{equation} \label{eq:current_density_expectation}
    \vec{{J}} := \tilde{\Tr} \{ \vec{j} \rho \} = \frac{1}{\vert\Omega\vert} \Tr \{ \vec{j} \rho \}, \quad \vec{j} := - \frac{e}{\hbar} \vec{\nabla} H.
\end{equation}
Thus, the current density in equilibrium is 
\begin{equation} \label{eq:J_eq}
    \vec{{J}}_{\text{eq}} := - \frac{e}{\hbar} \tilde{\Tr} \left\{ (\vec{\nabla} H) \Phi(H) \right\}.
\end{equation}

The equilibrium current \eqref{eq:J_eq} vanishes for free particles because of the invariance (even-ness) of the free dispersion relation under $\vec{k} \mapsto - \vec{k}$. The equilibrium current vanishes, similarly, for particles in a periodic potential, as long as the Bloch bands $E_n(\vec{k})$ are even under $\vec{k} \mapsto - \vec{k}$. A sufficient condition for this to hold is realness of the Hamiltonian $\overline{H} = H$ (often called ``time-reversal symmetry'' condition). This holds for tight-binding models with real coefficients, or continuum Hamiltonians $- \Delta + V$, with $V$ real, and hence holds for essentially all models of materials (e.g. graphene, twisted bilayer graphene) in the absence of an applied magnetic field. Alternatively, it suffices to have invariance of the Hamiltonian $H$ under $\vec{x} \mapsto - \vec{x}$ (often called ``parity symmetry''). For a discussion, see, for example, \cite{fefferman_weinstein_diracpoints}. 


\subsection{Result for constant applied field $\vec{{E}}$}

We now make our result precise for the case of time-independent (constant) $\vec{{E}}$. Consider the long-time average of the current density expectation $\vec{{J}}$ \eqref{eq:current_density_expectation}, averaged over scattering times modeled by \eqref{eq:poisson} 
\begin{equation} \label{eq:average_constant_case}
    \langle \vec{{J}} \rangle := \lim_{n \rightarrow \infty} \left\langle \frac{1}{t_n} \inty{0}{t_n}{ \vec{{J}}(t') }{t'} \right\rangle_{\Gamma}.
\end{equation}
More precisely, $\langle \cdot \rangle_\Gamma$ denotes taking the expectation over the Poisson distributions defining each $\tau_n := t_{n+1} - t_n$ \eqref{eq:poisson}. Then, write the vectors $\langle \vec{{J}} \rangle$, $\vec{{J}}_{\text{eq}}$, and $\vec{{E}}$, in terms of their components as
\begin{equation} \label{eq:components}
    \langle \vec{{J}} \rangle = \left( \langle {J}_l \rangle \right)_{1 \leq l \leq d}, \quad \vec{{J}}_{\text{eq}} = \left( {J}_{\text{eq},l} \right)_{1 \leq l \leq d}, \quad \vec{{E}} = \left( {E}_m \right)_{1 \leq m \leq d}.
\end{equation}
Our main result for constant $\vec{{E}}$ is the following expression for $\langle \vec{{J}} \rangle$ in terms of $\vec{{E}}$, known as the Kubo formula:
\begin{equation} \label{eq:DC_Kubo}
    \begin{split}
        &\langle {J}_l \rangle = {J}_{\text{eq},l} + \sum_{m = 1}^d \sigma_{l m} {E}_m + O\left( \vec{{E}}^2 \right), \quad 1 \leq l \leq d, \\
        &\sigma_{l m} := - \frac{e^2}{\hbar^2} \tilde{\Tr}\left\{ (\de_l H) \left( \mathcal{L}_H + \Gamma \right)^{-1} \de_m \Phi(H) \right\}, \quad 1 \leq l, m \leq d.
    \end{split}
\end{equation}
Note that \eqref{eq:DC_Kubo} agrees with \eqref{eq:standard_Kubo} when $\omega = 0$. We present a systematic formal derivation of \eqref{eq:DC_Kubo} in Section \ref{sec:DC_derivation}.

\subsection{Result for time-harmonic applied field $\vec{{E}}$}

To state our result precisely for the case of time-dependent $\vec{{E}}$ is somewhat more involved. Let $\vec{{E}}$ be time-harmonic with frequency $\omega \in \mathbb{R}$, so that
   \begin{equation} \label{eq:freq}\vec{{E}}(t;\theta) := \frac{1}{2 \pi} e^{- i (\omega t + \theta)} \vec{{E}}(\omega),
\end{equation}
where $\vec{E}(\omega)$ is a fixed vector, and $\theta \in [-\pi,\pi)$ is a parameter which we will average over to ease calculations; see \eqref{eq:time_average_timedep}. It seems possible that our results hold without this averaging step, especially since it is not necessary in the classical case (see \eqref{eq:classical_freq}), but we do not investigate this in this work. 
We define $\rho(t; \theta)$ and $\vec{J}(t; \theta)$ by \eqref{eq:vonNeumann_scattering} and \eqref{eq:current_density_expectation}, with $\vec{E}$ given by \eqref{eq:freq}.

We consider the following quantity
\begin{equation} \label{eq:time_average_timedep}
    \langle \vec{{J}} \rangle(\omega) := \lim_{n \rightarrow \infty} \left\langle \frac{1}{t_n} \inty{0}{t_n}{ \inty{- \pi}{\pi}{ e^{i (\omega t + \theta)} \vec{{J}}(t; \theta) }{\theta} }{t} \right\rangle_\Gamma.
\end{equation}
Note that we average over the phase $\theta$ of the applied field \eqref{eq:freq}. This simplifies the calculation considerably; see \eqref{eq:where_we_use_phase_averaging}. Note that this is unnecessary in the classical case, where the  Liouville equation (the classical counterpart of the von Neumann equation used here) can be solved explicitly in simple closed form even for time-dependent fields; see \eqref{eq:explicit_between}. 

We introduce notation for the vector components of $\langle \vec{{J}} \rangle$ and $\vec{{E}}$ as
\begin{equation} \label{eq:components_2}
    \langle \vec{{J}} \rangle(\omega) = \left( \vphantom{{E}_m} \langle {{J}}_l \rangle(\omega) \right)_{1 \leq l \leq d}, \quad \vec{{E}}(\omega) = \left( {{E}}_m(\omega) \right)_{1 \leq m \leq d}.
\end{equation}
Our main result for time-dependent $\vec{{E}}$ is given by the following expressions:
\begin{align} \label{eq:general_Kubo}
        &\langle {{J}}_l \rangle(\omega) = \sum_{m = 1}^d \sigma_{l m}(\omega) {{E}}_m(\omega) + O\left( \vec{{E}}^2 \right), \quad 1 \leq l \leq d,    \\ \nonumber
        &\sigma_{l m}(\omega) := - \frac{e^2}{\hbar^2} \tilde{\Tr} \left\{ (\de_l H) \left( \mathcal{L}_H - i \omega + \Gamma \right)^{-1} \de_m \Phi(H) \right\}, \quad 1 \leq l, m \leq d.
\end{align}
We present a systematic formal derivation of \eqref{eq:general_Kubo} in Section \ref{sec:AC_derivation}. In Section \ref{sec:reductions}, we study the limit of \eqref{eq:general_Kubo} as $\vert \Omega \vert \rightarrow \infty$. Therefore, we derive important reductions of \eqref{eq:general_Kubo} in special cases of the Hamiltonian $H$, namely, for free particles and particles in a periodic potential. 

\section{Derivation of general Kubo formula} \label{sec:quantum_derivation}

\subsection{Derivation for constant applied field $\vec{{E}}$} \label{sec:DC_derivation}

We start by re-writing the current density expectation averaged over time and scattering events \eqref{eq:average_constant_case} using a Tauberian theorem \cite{korevaar} as
\begin{equation} \label{eq:Tauberian}
    \langle \vec{{J}} \rangle = \lim_{n \rightarrow \infty} \left\langle \frac{1}{t_n} \inty{0}{t_n}{ \vec{{J}}(t') }{t'} \right\rangle_\Gamma = \lim_{\delta \downarrow 0} \lim_{n \rightarrow \infty} \left\langle \delta \inty{0}{t_n}{ e^{- \delta t'} \vec{{J}}(t') }{t'} \right\rangle_\Gamma.
\end{equation}
We can now split the infinite integral into integrals over intervals between scattering events
\begin{equation}
    \langle \vec{{J}} \rangle = \lim_{\delta \downarrow 0} \lim_{n \rightarrow \infty} \delta \sum_{m = 0}^{n-1} \left\langle \inty{t_{m}}{t_{m+1}}{ e^{- \delta t'} \vec{{J}}(t')  }{t'} \right\rangle_{\Gamma}.
\end{equation}
Substituting the formula for $\vec{{J}}$ \eqref{eq:current_density_expectation} and exchanging the orders of trace and time-integral and trace and summation over $n$, we have
\begin{equation} \label{eq:timeaveraged}
    \langle \vec{{J}} \rangle = - \frac{e}{\hbar} \lim_{\delta \downarrow 0} \lim_{n \rightarrow \infty} \delta \tilde{\Tr} \left\{ (\vec{\nabla} H) \sum_{m = 0}^{n-1} \left\langle \inty{t_m}{t_{m+1}}{ e^{- \delta t'} \rho(t') }{t'} \right\rangle_{\Gamma} \right\}.
\end{equation}
We now write $\rho$ in terms of the propagator for \eqref{eq:vonNeumann_scattering}
\begin{equation}
    \rho(t') = e^{- (t' - t_n) \mathcal{L}_{H_{\vec{{E}}}}} \Phi(H).
\end{equation}
We can then compute the integral
\begin{equation}
    \begin{split}
        &\inty{t_m}{t_{m+1}}{ e^{- \delta t'} \rho(t') }{t'} = \inty{t_m}{t_{m+1}}{ e^{- \delta t'} e^{- (t' - t_m) \mathcal{L}_{H_{\vec{{E}}}}}\Phi(H) }{t'} \\
        &= ( \mathcal{L}_{H_{\vec{{E}}}} + \delta )^{-1} e^{- \delta t_m} \left( I - e^{- (\mathcal{L}_{H_{\vec{{E}}}} + \delta ) \tau_m} \right) \Phi(H),
    \end{split}
\end{equation}
where $\tau_m := t_{m+1} - t_m$. Substituting this back into \eqref{eq:timeaveraged} we have
\begin{equation} \label{eq:timeaveraged_2}
    \begin{split}
        &\langle \vec{{J}} \rangle = - \frac{e}{\hbar} \lim_{\delta \downarrow 0} \lim_{n \rightarrow \infty} \delta   \\
        &\times \tilde{\Tr} \left\{ (\vec{\nabla} H) ( \mathcal{L}_{H_{\vec{{E}}}} + \delta )^{-1} \sum_{m = 0}^{n-1} \left\langle e^{- \delta t_m} \left( I - e^{- (\mathcal{L}_{H_{\vec{{E}}}} + \delta) \tau_m} \right) \right\rangle_{\Gamma} \Phi(H) \right\}.
    \end{split}
\end{equation}
We now perform the averaging. We will do so in detail for the $m = 0$ and $m = 1$ terms; the pattern for $m \geq 2$ will then be clear. The $m = 0$ term is (recall that $t_0 = 0$)
\begin{equation}
    \begin{split}
        \left\langle e^{- \delta t_0} \left( I - e^{- ( \mathcal{L}_{H_{\vec{{E}}}} + \delta ) \tau_0} \right) \right\rangle_{\Gamma} &= \inty{0}{\infty}{ \Gamma e^{- \Gamma \tau_0} \left( I - e^{- ( \mathcal{L}_{H_{\vec{{E}}}} + \delta ) \tau_0} \right) }{\tau_0}    \\
        &= I - \Gamma ( \mathcal{L}_{H_{\vec{{E}}}} + \delta + \Gamma )^{-1}.
    \end{split}
\end{equation}
The $m = 1$ term is
\begin{equation}
    \left\langle e^{- \delta t_1} \left( I - e^{- (\mathcal{L}_{H_{\vec{{E}}}} + \delta) \tau_1} \right) \right\rangle_{\Gamma} = \left\langle e^{- \delta \tau_0} \left( I - e^{- (\mathcal{L}_{H_{\vec{{E}}}} + \delta) \tau_1} \right) \right\rangle_{\Gamma},
\end{equation}
which we have to average over both $\tau_0$ and $\tau_1$. Since the variables are independent, we can simply average over each variable in turn. The average over $\tau_1$ obviously gives the same factor as before, and the average over $\tau_0$ gives
\begin{equation}
    \inty{0}{\infty}{ \Gamma e^{- \Gamma \tau_0} e^{- \delta \tau_0} }{\tau_0} = \frac{ \Gamma }{ \Gamma + \delta }.
\end{equation}
Hence, the $m = 1$ term in the sum is
\begin{equation}
        \left\langle e^{- \delta t_1} \left( I - e^{- (\mathcal{L}_{H_{\vec{{E}}}} + \delta) \tau_1} \right) \right\rangle_{\Gamma} = \left( \frac{ \Gamma }{ \Gamma + \delta } \right) \left( I - \Gamma ( \mathcal{L}_{H_{\vec{{E}}}} + \delta + \Gamma )^{-1} \right).
\end{equation}
Continuing in this way, using the fact that
\begin{equation}
    e^{- \delta t_m} = \prod_{j = 0}^{m-1} e^{- \delta \tau_j},
\end{equation}
we find that the $m$th term in the series is
\begin{equation}
    \left( \frac{ \Gamma }{ \Gamma + \delta } \right)^m \left( I - \Gamma ( \mathcal{L}_{H_{\vec{{E}}}} + \delta + \Gamma )^{-1} \right).
\end{equation}
We can sum the series, obtaining
\begin{equation} \label{eq:summed_series}
        \sum_{m = 0}^\infty \left\langle e^{- \delta t_m} \left( I - e^{- (\mathcal{L}_{H_{\vec{{E}}}} + \delta ) \tau_m} \right) \right\rangle_{\Gamma} = \frac{\Gamma + \delta }{ \delta } \left( I - \Gamma ( \mathcal{L}_{H_{\vec{{E}}}} + \delta + \Gamma )^{-1} \right).
\end{equation}
Substituting \eqref{eq:summed_series} into \eqref{eq:timeaveraged_2} we have
\begin{align} \label{eq:timeaveraged_3}
        &\langle \vec{{J}} \rangle =     \\ \nonumber
        &- \frac{e}{\hbar} \lim_{\delta \downarrow 0} (\Gamma + \delta) \tilde{\Tr} \left\{ (\vec{\nabla} H) ( \mathcal{L}_{H_{\vec{{E}}}} + \delta )^{-1} \left( I - \Gamma ( \mathcal{L}_{H_{\vec{{E}}}} + \delta + \Gamma )^{-1} \right) \Phi(H) \right\}.
\end{align}
The second resolvent identity (see, for example, Proposition 1.9 of \cite{HislopSigal}) essentially states that $A^{-1} - B^{-1} = A^{-1} (B - A) B^{-1}$ for invertible operators $A$ and $B$. Applying this identity with $A = \mathcal{L}_{H_{\vec{{E}}}} + \delta + \Gamma$ and $B = \mathcal{L}_{H_{\vec{E}}} + \delta$, we have
\begin{equation}
    \begin{split}
        &( \mathcal{L}_{H_{\vec{{E}}}} + \delta + \Gamma )^{-1}   \\
        &= ( \mathcal{L}_{H_{\vec{{E}}}} + \delta )^{-1} - \Gamma ( \mathcal{L}_{H_{\vec{{E}}}} + \delta )^{-1} ( \mathcal{L}_{H_{\vec{{E}}}} + \delta + \Gamma )^{-1}    \\ 
        &= ( \mathcal{L}_{H_{\vec{{E}}}} + \delta  )^{-1} \left( I  - \Gamma ( \mathcal{L}_{H_{\vec{{E}}}} + \delta + \Gamma )^{-1} \right),
    \end{split}
\end{equation}
so that \eqref{eq:timeaveraged_3} simplifies to
\begin{equation} \label{eq:timeaveraged_32}
    \langle \vec{{J}} \rangle = - \frac{e}{\hbar} \lim_{\delta \downarrow 0} (\Gamma + \delta ) \tilde{\Tr} \left\{ (\vec{\nabla} H) ( \mathcal{L}_{H_{\vec{{E}}}} + \delta + \Gamma)^{-1} \Phi(H) \right\}.
\end{equation}
Exchanging the order of taking the trace and taking the limit $\delta \downarrow 0$ we arrive at
\begin{equation} \label{eq:timeaveraged_33}
    \langle \vec{{J}} \rangle = - \frac{e}{\hbar} \tilde{\Tr} \left\{ \lim_{\delta \downarrow 0} (\Gamma + \delta) (\vec{\nabla} H) ( \mathcal{L}_{H_{\vec{{E}}}} + \delta + \Gamma)^{-1} \Phi(H) \right\},
\end{equation}
where it is trivial to take the limit, resulting in
\begin{equation} \label{eq:timeaveraged_4}
    \langle \vec{{J}} \rangle = - \frac{e \Gamma}{\hbar} \tilde{\Tr} \left\{ (\vec{\nabla} H) ( \mathcal{L}_{H_{\vec{{E}}}} + \Gamma )^{-1} \Phi(H) \right\}.
\end{equation}
We now note that
\begin{equation} \label{eq:resolvent_id}
    \begin{split}
        ( \mathcal{L}_{H_{\vec{{E}}}} + \Gamma )^{-1} &= ( \mathcal{L}_{H} + \Gamma )^{-1} + \left( ( \mathcal{L}_{H_{\vec{{E}}}} + \Gamma )^{-1} - ( \mathcal{L}_{H} + \Gamma )^{-1} \right)   \\
        &= ( \mathcal{L}_{H} + \Gamma )^{-1} + ( \mathcal{L}_{H_{\vec{{E}}}} + \Gamma )^{-1} \left( \frac{e}{\hbar} \vec{{E}} \cdot \vec{\nabla} \right) ( \mathcal{L}_H + \Gamma )^{-1},
    \end{split} 
\end{equation}
where the second equality follows from 
\begin{equation} \label{eq:perturbation}
    \mathcal{L}_{H_{\vec{{E}}}} = \mathcal{L}_H - \frac{e}{\hbar} \vec{{E}} \cdot \vec{\nabla}
\end{equation}
and the second resolvent identity~\cite{HislopSigal}. Substituting \eqref{eq:resolvent_id} into \eqref{eq:timeaveraged_4} separates \eqref{eq:timeaveraged_4} into equilibrium (corresponding to the first term on the right-hand side of \eqref{eq:resolvent_id}) and non-equilibrium (corresponding to the second term) contributions to the conductivity. Hence, noting that $( \mathcal{L}_H + \Gamma )^{-1} \Phi(H) = \Gamma^{-1} \Phi(H)$, we simplify \eqref{eq:timeaveraged_4} to 
\begin{equation} \label{eq:timeaveraged_5}
    \langle \vec{{J}} \rangle = \vec{{J}}_{\text{eq}} - \frac{e \Gamma}{\hbar} \tilde{\Tr} \left\{ (\vec{\nabla} H) ( \mathcal{L}_{H_{\vec{{E}}}} + \Gamma )^{-1} \left( \frac{e}{\hbar} \vec{{E}} \cdot \vec{\nabla} \right) ( \mathcal{L}_H + \Gamma )^{-1} \Phi(H) \right\},
\end{equation}
which further simplifies to
\begin{equation} \label{eq:timeaveraged_6}
    \langle \vec{{J}} \rangle = \vec{{J}}_{\text{eq}} - \frac{e^2}{\hbar^2} \tilde{\Tr} \left\{ (\vec{\nabla} H) ( \mathcal{L}_{H_{\vec{{E}}}} + \Gamma )^{-1} \left( \vec{{E}} \cdot \vec{\nabla} \right) \Phi(H) \right\}.
\end{equation}
It is worth emphasizing that the calculation up to this point has been exact, in the sense that we have not made any expansion in powers of $\vec{E}$. We now make such an approximation. Using \eqref{eq:perturbation} and the second resolvent identity once more, we obtain
\begin{equation}
    ( \mathcal{L}_{H_{\vec{{E}}}} + \Gamma )^{-1} = ( \mathcal{L}_{H} + \Gamma )^{-1} + O( \vec{{E}} ).
\end{equation}
Substituting this into \eqref{eq:timeaveraged_6}, and writing everything out in components, we obtain \eqref{eq:DC_Kubo}. 

\subsection{Derivation for time-harmonic applied field $\vec{{E}}$} \label{sec:AC_derivation}

The first task is to find a convenient representation of the propagator for the von Neumann equation \eqref{eq:vonNeumann_scattering} when $\vec{{E}}$ is given by \eqref{eq:freq} and thus time-dependent. It is convenient to introduce notation
\begin{equation} \label{eq:mathcal_regular_E}
    \vec{\mathcal{E}}(\theta') := \frac{1}{2 \pi} e^{- i \theta'} {\vec{E}}(\omega),  
\end{equation}
so that $\vec{{E}}(t; \theta) = \vec{\mathcal{E}}(\omega t + \theta)$ and the von Neumann equation is
\begin{equation} \label{eq:vonNeumann_scattering2}
    \de_{t} \rho(t,\theta)  = \left( - \mathcal{L}_H + \frac{e}{\hbar} \vec{\mathcal{E}}(\omega t + \theta) \cdot \vec{\nabla} \right) \rho(t,\theta), \quad \rho(t_n,\theta) = \Phi(H).
\end{equation}
We introduce a change of variables $t'=t$ and $\theta' = \theta + \omega t$ to \eqref{eq:vonNeumann_scattering2} and  a new function $P$ such that  $\rho(t;\theta) =  P(t',\theta')$ 
 which evolves by the equivalent PDE
\begin{equation} \label{eq:hypeq}
   \begin{split}
        &\de_{t'} P(t',\theta') = \left( - \mathcal{L}_H + \frac{e}{\hbar} \vec{\mathcal{E}}(\theta') \cdot \vec{\nabla} - \omega \de_{\theta'} \right) P(t',\theta') ,\\\
        &P(t_n,\theta') = \Phi(H).
   \end{split}
\end{equation}
For each $\theta'$, this equation has constant coefficients in $t'$, so we can write its solution as
\begin{equation} \label{eq:simple_propagator}
    P(t',\theta') = e^{- (t' - t_n) \left( \mathcal{L}_H - \frac{e}{\hbar} \vec{\mathcal{E}}(\theta') \cdot \vec{\nabla} + \omega \de_{\theta'} \right)} \Phi(H). 
\end{equation}
We note that the von Neumann equation \eqref{eq:vonNeumann_scattering} is the characteristic equation for \eqref{eq:hypeq} and that $(t,\theta)$ are the characteristic variables.

We now again use a Tauberian theorem to write
\begin{equation} \label{eq:average_general_2}
    \langle \vec{{J}} \rangle(\omega) = \lim_{\delta \downarrow 0} \lim_{n \rightarrow \infty} \left\langle \delta \inty{0}{t_n}{ \inty{-\pi}{\pi}{ e^{- \delta t} e^{i (\omega t + \theta)} \vec{{J}}(t;\theta) }{\theta} }{t} \right\rangle_{\Gamma}.
\end{equation}
Splitting up the time integral and substituting the forms of $\vec{{J}}$ and $\rho$ we have
\begin{equation*}
    \langle \vec{{J}} \rangle(\omega) = - \frac{e}{\hbar} \lim_{\delta \downarrow 0} \lim_{n \rightarrow \infty} \tilde{\Tr} \left\{ (\vec{\nabla} H) \left\langle \delta \sum_{m = 0}^{n-1} \inty{t_m}{t_{m+1}}{ \inty{-\pi}{\pi}{ e^{- \delta t} e^{i (\omega t + \theta)} P(t,\theta + \omega t) }{\theta} }{t} \right\rangle_{\Gamma} \right\}.
\end{equation*}
Changing variable in the $\theta$ integral to $\theta' = \theta + \omega t$ we find
\begin{equation} \label{eq:where_we_use_phase_averaging}
    \langle \vec{{J}} \rangle(\omega) = - \frac{e}{\hbar} \lim_{\delta \downarrow 0} \lim_{n \rightarrow \infty} \tilde{\Tr} \left\{ (\vec{\nabla} H) \left\langle \delta \sum_{m = 0}^{n-1} \inty{t_m}{t_{m+1}}{ \inty{-\pi}{\pi}{ e^{- \delta t} e^{i \theta} P(t,\theta) }{\theta} }{t} \right\rangle_{\Gamma} \right\}.
\end{equation}
Substituting the form of $P$ we have
\begin{align}
        &\langle \vec{{J}} \rangle(\omega) = - \frac{e}{\hbar} \lim_{\delta \downarrow 0} \lim_{n \rightarrow \infty} \tilde{\Tr} \left\{ \vphantom{\int_{t_n}^{t_{n+1}}} (\vec{\nabla} H) \delta \right. \\ \nonumber
        &\left. \times \left\langle \sum_{m = 0}^{n-1} \inty{t_m}{t_{m+1}}{ \inty{-\pi}{\pi}{ e^{- \delta t} e^{i \theta} e^{- (t - t_m) \left( \mathcal{L}_H - \frac{e}{\hbar} \vec{\mathcal{E}}(\theta) \cdot \vec{\nabla} + \omega \de_{\theta} \right)} \Phi(H) }{\theta} }{t} \right\rangle_{\Gamma} \right\},
\end{align}
which we can re-write as
\begin{align}
        &\langle \vec{{J}} \rangle(\omega) = - \frac{e}{\hbar} \lim_{\delta \downarrow 0} \lim_{n \rightarrow \infty} \tilde{\Tr} \left\{ \vphantom{\int_{t_m}^{t_{m+1}}} (\vec{\nabla} H) \delta \right. \\ \nonumber
        &\left. \times \left\langle \sum_{m = 0}^{n-1} \inty{t_m}{t_{m+1}}{ \inty{-\pi}{\pi}{ e^{- \delta t_m} e^{i \theta} e^{- (t - t_m) \left( \mathcal{L}_H - \frac{e}{\hbar} \vec{\mathcal{E}}(\theta) \cdot \vec{\nabla} + \omega \de_{\theta} + \delta \right)} \Phi(H) }{\theta} }{t} \right\rangle_{\Gamma} \right\}.
\end{align}
Exchanging the orders of the $\delta$ and $n$ limits with the trace, integral over $\theta$, and average over the Poisson process, we find
\begin{align}
        &\langle \vec{{J}} \rangle(\omega) = - \frac{e}{\hbar} \tilde{\Tr} \left\{ \vphantom{\int_{t_n}^{t_{n+1}}} (\vec{\nabla} H) \right. \\ \nonumber
        &\left. \times \inty{-\pi}{\pi}{ e^{i \theta} \lim_{\delta \downarrow 0} \lim_{n \rightarrow \infty} \delta \left\langle \sum_{m = 0}^{n-1} e^{- \delta t_m} \inty{t_m}{t_{m+1}}{ e^{- (t - t_m) \left( \mathcal{L}_H - \frac{e}{\hbar} \vec{\mathcal{E}}(\theta) \cdot \vec{\nabla} + \omega \de_{\theta} + \delta \right)} \Phi(H) }{t} \right\rangle_{\Gamma} }{\theta} \right\}.
\end{align}
We are now back to the setting of the DC case. By the same steps: performing the $t$ integrals, averaging over the Poisson process, summing up the series, and then manipulating using the second resolvent identity~\cite{HislopSigal}, we arrive at
\begin{equation}
    \begin{split}
        &\langle \vec{{J}} \rangle(\omega) = - \frac{e}{\hbar} \tilde{\Tr} \left\{ \vphantom{\int_{t_n}^{t_{n+1}}} (\vec{\nabla} H) \right. \\
        &\left. \times \inty{-\pi}{\pi}{ e^{i \theta} \lim_{\delta \downarrow 0} (\Gamma + \delta) \left( \mathcal{L}_H - \frac{e}{\hbar} \vec{\mathcal{E}}(\theta) \cdot \vec{\nabla} + \omega \de_{\theta} + \delta + \Gamma \right)^{-1} \Phi(H) }{\theta} \right\}.
    \end{split}
\end{equation}
Taking the limit then yields
\begin{equation*}
    \langle \vec{{J}} \rangle(\omega) = - \frac{e \Gamma}{\hbar} \tilde{\Tr}\left\{ \vphantom{\int_{t_n}^{t_{n+1}}} (\vec{\nabla} H) \inty{-\pi}{\pi}{ e^{i \theta} \left( \mathcal{L}_H - \frac{e}{\hbar} \vec{\mathcal{E}}(\theta) \cdot \vec{\nabla} + \omega \de_{\theta} + \Gamma \right)^{-1} \Phi(H) }{\theta} \right\}.
\end{equation*}
Just as in the DC case, we can use the second resolvent identity to isolate the non-equilibrium part, (the equilibrium part contributes $0$ for $\omega \neq 0$ because of the integral over $\theta$), and then using $(\mathcal{L}_H + \omega \de_\theta + \Gamma)^{-1} \Phi(H) = \Gamma^{-1} \Phi(H)$, we arrive at
\begin{equation*}
    \langle \vec{{J}} \rangle(\omega) = - \frac{e^2}{\hbar^2} \tilde{\Tr} \left\{ \vphantom{\int_{t_n}^{t_{n+1}}} (\vec{\nabla} H) \inty{-\pi}{\pi}{ e^{i \theta} \left( \mathcal{L}_{H_{\vec{\mathcal{E}}}} + \omega \de_{\theta} + \Gamma \right)^{-1} \vec{\mathcal{E}}(\theta) }{\theta} \cdot \vec{\nabla} \Phi(H) \right\}.
\end{equation*}
Again, we emphasize that the calculation is, up to this point, exact with respect to $\vec{E}$. Just as in the DC case, we now use the second resolvent identity to take the limit $\vec{\mathcal{E}} \rightarrow 0$ in the resolvent. We can then insert the formula for $\vec{\mathcal{E}}$ \eqref{eq:mathcal_regular_E} and evaluate the integral over $\theta$ to obtain 
\begin{equation}
    \langle \vec{{J}} \rangle(\omega) = - \frac{e^2}{\hbar^2} \tilde{\Tr} \left\{ \vphantom{\int_{t_n}^{t_{n+1}}} (\vec{\nabla} H) \left( \mathcal{L}_H - i \omega + \Gamma \right)^{-1} \vec{{E}}(\omega) \cdot \vec{\nabla} \Phi(H) \right\} + O(\vec{{E}}^2),
\end{equation}
from which we obtain \eqref{eq:general_Kubo}. 

\section{Applications of the Kubo formula} \label{sec:reductions}

\subsection{Free particles} \label{sec:free}

We consider the special case of free particles, taking $\Omega$ to be a $d$-dimensional box with sides of length $L$, so that $\vert\Omega\vert = L^d$. The Hilbert space, $\mathcal{H}$, and Hamiltonian, $H$, are
\begin{equation}
    \mathcal{H} = L^2\left(\left[-\frac{L}{2},\frac{L}{2}\right]^d\right), \quad H = - \frac{\hbar^2}{2 m} \Delta_{\vec{x}}.
\end{equation}
It is now straightforward to compute the current operator (recall \eqref{eq:derivation} and \eqref{eq:current_density_expectation})
\begin{equation} \label{eq:free_current}
    \vec{j} = - \frac{e i}{\hbar} \left[\vec{X}, \frac{\hbar^2 \Delta_{\vec{x}}}{2 m} \right] = - \frac{e \hbar}{m} (- i \vec{\nabla}_{\vec{x}}).
\end{equation}
Note that we adopt the notation of putting a subscript $\vec{x}$ to denote ordinary differential operators in order to distinguish from our notation for derivations \eqref{eq:derivation}. A natural orthonormal basis for $\mathcal{H}$ is given by the Fourier modes
\begin{equation}
    \left\{ \frac{e^{i \vec{k} \cdot \vec{x}}}{L^{\frac{d}{2}}} : \vec{k} \in \frac{ 2 \pi }{ L } \mathbb{Z}^d \right\}.
\end{equation}
The particle number $N$ is given by the trace of the Fermi-Dirac distribution over $\mathcal{H}$, which we can evaluate as 
\begin{equation}
    N := \Tr \left\{ e^{\beta(H - \mu)} + 1 \right\}^{-1} = \sum_{\vec{k} \in \frac{2 \pi}{L} \mathbb{Z}^d} \left\{ e^{\beta \left(\frac{\hbar^2 \vert\vec{k}\vert^2}{2 m} - \mu\right)} + 1 \right\}^{-1}.
\end{equation}
Taking the limit $L \rightarrow \infty$, the sum converges to an integral, and we derive the relationship between particle density $\tilde{N}$ and chemical potential $\mu$
\begin{equation} \label{eq:density_0}
    \tilde{N} := \lim_{L \rightarrow \infty} \frac{N}{L^d} = \frac{ 1 }{ (2 \pi)^d } \inty{\mathbb{R}^d}{}{ \left\{ e^{\beta \left( \frac{\hbar^2 \vert \vec{k} \vert^2}{2 m} - \mu \right)} + 1 \right\}^{-1} }{\vec{k}}.
\end{equation}
It is straightforward to verify that the current density vanishes in equilibrium:
\begin{equation}
    \frac{1}{L^d} \Tr \{ \vec{j} \Phi(H) \} = - \frac{e \hbar}{m L^d} \sum_{\vec{k} \in \frac{2 \pi}{L} \mathbb{Z}^d} \vec{k} \left\{ e^{ \beta \left( \frac{\hbar^2 \vert \vec{k} \vert^2}{2 m} - \mu \right) } + 1 \right\}^{-1} = 0,
\end{equation}
because of even-ness of the Fermi-Dirac distribution with respect to $\vec{k}$.

Evaluating the trace in the Kubo formula \eqref{eq:general_Kubo}, we derive
\begin{align}
        &\sigma_{lm} (\omega) = - \frac{e^2}{\hbar^2 L^d} \\ \nonumber \times  
        &\sum_{\vec{k} \in \frac{2 \pi}{L} \mathbb{Z}^d} \left( \frac{\hbar^2}{m} k_l \right) ( \Gamma - i \omega )^{-1} \left( - \frac{\beta \hbar^2 k_m}{m} \right) e^{ \beta \left(\frac{\hbar^2 \vert \vec{k} \vert^2}{2 m} - \mu\right) } \left\{ e^{ \beta \left(\frac{\hbar^2 \vert \vec{k} \vert^2}{2 m} - \mu\right) } + 1 \right\}^{-2}.
\end{align}
Anti-symmetry in $k$ implies immediately that $\sigma_{lm}(\omega) = 0$ unless $l = m$, so we derive, after taking $L \rightarrow \infty$,
\begin{equation}
    \sigma_{lm} (\omega) = - \delta_{lm} \frac{e^2}{m (2 \pi)^d ( \Gamma - i \omega )} \inty{\mathbb{R}^d}{}{ k_l \pdf{k_l} \left\{ e^{ \beta \left( \frac{\hbar^2 \vert \vec{k} \vert^2}{2 m} - \mu \right) } + 1 \right\}^{-1} }{\vec{k}}.
\end{equation}
Integrating by parts, we arrive at (recall \eqref{eq:density_0})
\begin{equation}
    \sigma_{lm} (\omega) = \delta_{lm} \frac{e^2}{m ( \Gamma - i \omega )} \tilde{N}, \quad \tilde{N} = \frac{N}{L^d},
\end{equation}
which is consistent with the literature; see, for example, \cite{ashcroft_mermin}.

\subsection{Particles in continuum periodic potential} \label{sec:periodic}

We now consider the case of particles in a periodic potential. 
Let $A$ denote a real and invertible $d \times d$ matrix. Then, we can introduce the Bravais lattice and real space unit cell
\begin{equation}
    \Lambda := \left\{ \vec{R} = A \vec{m} : \vec{m} \in \mathbb{Z}^d \right\}, \quad \Gamma := A \left[-\frac{1}{2},\frac{1}{2}\right)^d,
\end{equation}
and the reciprocal lattice and momentum space unit cell (Brillouin zone)
\begin{equation}
    \Lambda^* := \left\{ \vec{G} = B \vec{n} : \vec{n} \in \mathbb{Z}^d \right\}, \quad \Gamma^* := B \left[-\frac{1}{2},\frac{1}{2}\right)^d, \quad B := 2 \pi A^{- T}.
\end{equation}
We then set
\begin{equation}
    \Omega = \Gamma_L := A \left[ - \frac{L}{2} , \frac{L}{2} \right)^d,
\end{equation}
so that $\vert\Omega\vert = \vert\Gamma\vert L^d$, where we restrict $L$ to the positive even integers. We consider the Hilbert space, $\mathcal{H}$, and Hamiltonian, $H$, given by
\begin{equation}
    \mathcal{H} := L^2\left( A \left[ - \frac{L}{2} , \frac{L}{2} \right)^d \right), \quad H := - \frac{\hbar^2}{2 m} \Delta_{\vec{x}} + V(\vec{x}),
\end{equation}
where we assume $V \in C^\infty(\mathbb{R}^d,\mathbb{R})$, and
\begin{equation}
    V(\vec{x} + \vec{R}) = V(\vec{x}), \quad \vec{x} \in \mathbb{R}^d, \vec{R} \in \Lambda.
\end{equation}
The current operator is, again, given by \eqref{eq:free_current}.

It is natural to introduce the basis of Bloch functions, defined as follows. Consider, for $\vec{k} \in \Gamma^*$, the eigenvalue problem defined by
\begin{equation}
    H \phi = E \phi, \quad \phi(\vec{x} + \vec{R};\vec{k}) = e^{i \vec{k} \cdot \vec{R}} \phi(\vec{x};\vec{k}), \quad \vec{x} \in \Gamma, \vec{R} \in \Lambda,
\end{equation}
for each $\vec{k} \in \Gamma^*$. The operator $H$ has compact resolvent for each $\vec{k}$, and hence a sequence of discrete eigenvalues which can be labelled with multiplicity
\begin{equation}
    E_1(\vec{k}) \leq E_2(\vec{k}) \leq ... \leq E_n(\vec{k}) \leq ...,
\end{equation}
with associated eigenfunctions $\phi_n(\vec{x};\vec{k})$. To be consistent with the free case, we assume that these functions are normalized in $L^2\left(\Gamma\right)$, i.e.,
\begin{equation}
    \inty{\Gamma}{}{ \vert\phi_n(\vec{x};\vec{k})\vert^2 }{\vec{x}} = 1, \quad n \in \mathbb{N}_{> 0}, \vec{k} \in \Gamma^*.
\end{equation}
Introduce the discretized Brillouin zone
\begin{equation}
    \Gamma_L^* := \left\{ \vec{k} = \frac{B \vec{n}}{L} : \vec{n} \in \left\{-\frac{L}{2},...,\frac{L}{2}-1\right\}^d \right\},
\end{equation}
then the set of Bloch functions
\begin{equation} \label{eq:Bloch_basis_discretized}
    \left\{ \frac{\phi_n(\cdot;\vec{k})}{L^{\frac{d}{2}}} : \vec{k} \in \Gamma_L^*, n \in \mathbb{N}_{> 0} \right\}
\end{equation}
forms an orthonormal basis of $L^2\left( A\left[-\frac{L}{2},\frac{L}{2}\right)^d \right)$. It is straightforward to check that each Bloch function can be decomposed as
\begin{equation}
    \phi_n(\vec{x};\vec{k}) = e^{i \vec{k} \cdot \vec{x}} \chi_n(\vec{x};\vec{k}),
\end{equation}
where the $\chi$'s satisfy the eigenvalue problem
\begin{gather}
    H(\vec{k}) \chi_n = E \chi_n, \quad H(\vec{k}) := \frac{\hbar^2}{2 m} ( \vec{k} - i \vec{\nabla}_{\vec{x}} )^2 + V(\vec{x}) \label{eq:H_k} \\ 
        \chi_n(\vec{x} + \vec{R};\vec{k}) = \chi_n(\vec{x};\vec{k}), \quad \vec{x} \in \Gamma, \vec{R} \in \Lambda,
\end{gather}
for each $\vec{k} \in \Gamma^*$.

It is now natural to evaluate the trace in the basis \eqref{eq:Bloch_basis_discretized} as
\begin{equation}
    N = \Tr \left\{ e^{\beta (H - \mu)} + 1 \right\}^{-1} = \sum_{n \in \mathbb{N}_{>0}} \sum_{\vec{k} \in \Gamma^*_L} \left\{ e^{\beta \left( E_n(\vec{k}) - \mu \right)} + 1 \right\}^{-1},
\end{equation}
where the sum over $n$ converges because of the Weyl asymptotics \cite{122017020120101}, which state that there exist positive constants $N, c_1, c_2$ such that
\begin{equation}
    c_1 n^{\frac{2}{d}} \leq E_n \leq c_2 n^{\frac{2}{d}}, \quad \forall n \geq N.
\end{equation}
Taking the limit $L \rightarrow \infty$, the sum converges to an integral and we obtain
\begin{equation}
    \tilde{N} := \lim_{L \rightarrow \infty} \frac{N}{L^d \vert\Gamma\vert} = \frac{1}{(2 \pi)^d} \sum_{n \in \mathbb{N}_{>0}} \inty{\Gamma^*}{}{ \left\{ e^{\beta \left( E_n(\vec{k}) - \mu \right)} + 1 \right\}^{-1} }{\vec{k}}.
\end{equation}

We now want to evaluate the trace formula with respect to the normalized Bloch basis \eqref{eq:Bloch_basis_discretized}. We start by considering the matrix elements
\begin{equation}
    \begin{split}
        &\frac{1}{L^d} \inty{\Gamma_L}{}{ \overline{ \phi_n(\vec{x};\vec{k}) } \left( \de_l H \right) \phi_{n'}(\vec{x};\vec{k}') }{\vec{x}}   \\
        &= \frac{ \hbar^2 }{ m L^d } \inty{\Gamma_L}{}{ \overline{ \phi_n(\vec{x};\vec{k}) } \left( - i \pdf{x_l} \right) \phi_{n'}(\vec{x};\vec{k}') }{\vec{x}}.
    \end{split}
\end{equation}
Expanding the Bloch functions we have
\begin{equation} \label{eq:sim}
    = \frac{ \hbar^2 }{ m L^d } \inty{\Gamma_L}{}{ e^{i ( \vec{k}' - \vec{k} ) \cdot \vec{x}} \overline{ \chi_n(\vec{x};\vec{k}) } \left(k_l - i \pdf{x_l}\right) \chi_{n'}(\vec{x};\vec{k}') }{\vec{x}}.
\end{equation}
Introducing the truncated lattice
\begin{equation}
    \Lambda_L := \left\{ \vec{R} = A \vec{m} : \vec{m} \in \left\{ - \frac{L}{2} , ... , \frac{L}{2} - 1 \right\}^d \right\},
\end{equation}
we can further simplify \eqref{eq:sim} as 
\begin{equation*}
    = \frac{ \hbar^2 }{ m L^d } \sum_{\vec{R} \in \Lambda_L} e^{i ( \vec{k}' - \vec{k} ) \cdot \vec{R}} \inty{\Gamma}{}{ e^{i ( \vec{k}' - \vec{k} ) \cdot \vec{x}} \overline{ \chi_n(\vec{x};\vec{k}) } \left(k_l - i \pdf{x_l}\right) \chi_{n'}(\vec{x};\vec{k}') }{\vec{x}}.
\end{equation*}
Evaluating the sum we find that 
\begin{equation}
    \frac{1}{L^d} \sum_{\vec{R} \in \Lambda_L} e^{i ( \vec{k}' - \vec{k} ) \cdot \vec{R}} = \delta_{\vec{k}\vec{k}'}, \quad \delta_{\vec{k}\vec{k}'} := \begin{cases} 1 & \vec{k} = \vec{k}', \\ 0 & \vec{k} \neq \vec{k}'. \end{cases}
\end{equation}
We thus have that
\begin{equation}
    \begin{split}
        &\frac{1}{L^d} \inty{\Gamma_L}{}{ \overline{ \phi_n(\vec{x};\vec{k}) } \left( \de_l H \right) \phi_{n'}(\vec{x};\vec{k}') }{\vec{x}}   \\
        &= \delta_{\vec{k}\vec{k}'} \frac{ \hbar^2 }{ m } \inty{\Gamma}{}{ \overline{ \chi_n(\vec{x};\vec{k}) } \left(k_l - i \pdf{x_l}\right) \chi_{n'}(\vec{x};\vec{k}) }{\vec{x}},
    \end{split}
\end{equation}
which can also be written succinctly in terms of $H(\vec{k})$ \eqref{eq:H_k} as
\begin{equation} \label{eq:succinct}
    \begin{split}
        &\frac{1}{L^d} \inty{\Gamma_L}{}{ \overline{ \phi_n(\vec{x};\vec{k}) } \left( \de_l H \right) \phi_{n'}(\vec{x};\vec{k}') }{\vec{x}}   \\
        &= \delta_{\vec{k}\vec{k}'} \inty{\Gamma}{}{ \overline{ \chi_n(\vec{x};\vec{k}) } \left( \pd{H}{k_l}(\vec{k}) \right) \chi_{n'}(\vec{x};\vec{k}) }{\vec{x}}.
    \end{split}
\end{equation}
We now simplify
\begin{align}
        &\inty{\Gamma_L}{}{ \overline{ \phi_n(\vec{x};\vec{k}) } \left( \mathcal{L}_H - i \omega + \Gamma \right)^{-1} \left( \de_m \Phi(H) \right) \phi_{n'}(\vec{x};\vec{k}') }{\vec{x}}   \\ \nonumber
        &= \frac{1}{ \frac{i}{\hbar} \left( E_n(\vec{k}) - E_{n'}(\vec{k}') \right) - i \omega + \Gamma } \inty{\Gamma_L}{}{ \overline{ \phi_n(\vec{x};\vec{k}) } \left( \de_m \Phi(H) \right) \phi_{n'}(\vec{x};\vec{k}') }{\vec{x}},
\end{align}
which can be further simplified as
\begin{equation}
    \begin{split}
        &\inty{\Gamma_L}{}{ \overline{ \phi_n(\vec{x};\vec{k}) } \left( \de_m \Phi(H) \right) \phi_{n'}(\vec{x};\vec{k}') }{\vec{x}}    \\
        &= - i \inty{\Gamma_L}{}{ \overline{ \phi_n(\vec{x};\vec{k}) } \left( x_m \Phi(H) - \Phi(H) x_m \right) \phi_{n'}(\vec{x};\vec{k}') }{\vec{x}}    \\
        &= \frac{ \Phi(E_{n'}(\vec{k}')) - \Phi(E_n(\vec{k})) }{ E_{n'}(\vec{k}') - E_n(\vec{k}) } \inty{\Gamma_L}{}{ \overline{ \phi_n(\vec{x};\vec{k}) } ( \de_m H ) \phi_{n'}(\vec{x};\vec{k}') }{\vec{x}}.
    \end{split}
\end{equation}
The above expression requires interpretation when $\vec{k} = \vec{k}'$ and $n = n'$, or in the presence of degeneracies; see \eqref{eq:Drude}. Using \eqref{eq:succinct}, we then have that
\begin{equation}
    \begin{split}
        &\frac{1}{L^d} \inty{\Gamma_L}{}{ \overline{ \phi_n(\vec{x};\vec{k}) } \left( \mathcal{L}_H - i \omega + \Gamma \right)^{-1} \left( \de_m \Phi(H) \right) \phi_{n'}(\vec{x};\vec{k}') }{\vec{x}}   \\
        &= \delta_{\vec{k}\vec{k}'} \frac{\left( \Phi( E_{n'}(\vec{k}) ) - \Phi( E_n(\vec{k}) ) \right) \inty{\Gamma}{}{ \overline{ \chi_n(\vec{x};\vec{k}) } \left( \pd{H}{k_m}(\vec{k}) \right) \chi_{n'}(\vec{x};\vec{k}) }{\vec{x}} }{ \left( E_{n'}(\vec{k}) - E_n(\vec{k}) \right) \left( \frac{i}{\hbar} \left( E_n(\vec{k}) - E_{n'}(\vec{k}) \right) - i \omega + \Gamma \right) }.
    \end{split}
\end{equation}
We can now evaluate the trace per unit volume as
\begin{align} \label{eq:periodic_Kubo}
        &\sigma_{lm}(\omega) = - \frac{e^2}{\hbar^2 \vert\Gamma\vert L^d} \sum_{n,n' \in \mathbb{N}_{> 0}} \sum_{\vec{k} \in \Gamma^*_L}   \\ \nonumber
        &\frac{ \left( \Phi(E_{n'}(\vec{k})) - \Phi(E_{n}(\vec{k})) \right) \ip{ \chi_{n}(\vec{k}) }{ \left( \pd{H}{k_l}(\vec{k}) \right) \chi_{n'}(\vec{k}) } \ip{ \chi_{n'}(\vec{k}) }{ \left( \pd{H}{k_m}(\vec{k}) \right) \chi_n(\vec{k}) } }{(E_{n'}(\vec{k}) - E_{n}(\vec{k}))\left(\frac{i}{\hbar} \left( E_{n'}(\vec{k}) - E_n(\vec{k}) \right) - i \omega + \Gamma\right)}.
\end{align}
Taking the limit $L \rightarrow \infty$, we obtain
\begin{align} \label{eq:general_periodic}
        &\sigma_{lm}(\omega) = - \frac{e^2}{\hbar^2 (2 \pi)^d} \sum_{n,n' \in \mathbb{N}_{> 0}} \inty{\Gamma^*}{}{}{\vec{k}}  \\ \nonumber
        &\frac{ \left( \Phi(E_{n'}(\vec{k})) - \Phi(E_{n}(\vec{k})) \right) \ip{ \chi_{n}(\vec{k}) }{ \left( \pd{H}{k_l}(\vec{k}) \right) \chi_{n'}(\vec{k}) } \ip{ \chi_{n'}(\vec{k}) }{ \left( \pd{H}{k_m}(\vec{k}) \right) \chi_n(\vec{k}) } }{(E_{n'}(\vec{k}) - E_{n}(\vec{k}))\left(\frac{i}{\hbar} \left( E_{n'}(\vec{k}) - E_n(\vec{k}) \right) - i \omega + \Gamma\right)},
\end{align}
which is consistent with the literature; see, for example, Chapter 13 of \cite{ashcroft_mermin}. It is often convenient to separate the $n = n'$ and $n \neq n'$ terms in this sum. The $n = n'$ terms, known as the Drude conductivity, are
\begin{equation} \label{eq:Drude}
    \begin{split}
        &\sigma^{\text{D}}_{lm}(\omega) = - \frac{e^2}{\hbar^2 (2 \pi)^d (\Gamma - i \omega)} \sum_{n \in \mathbb{N}_{> 0}} \inty{\Gamma^*}{}{}{\vec{k}}  \\
        &\fd{\Phi}{E}(E_n(\vec{k})) \ip{ \chi_{n}(\vec{k}) }{ \left( \pd{H}{k_l}(\vec{k}) \right) \chi_{n}(\vec{k}) } \ip{ \chi_{n}(\vec{k}) }{ \left( \pd{H}{k_m}(\vec{k}) \right) \chi_n(\vec{k}) }.
    \end{split}
\end{equation}
If every band of $H(\vec{k})$ is simple, the Drude conductivity can be simplified further as
\begin{equation*}
    \sigma^{\text{D}}_{lm}(\omega) = - \frac{e^2}{\hbar^2 (2 \pi)^d (\Gamma - i \omega)} \sum_{n \in \mathbb{N}_{> 0}} \inty{\Gamma^*}{}{ \fd{\Phi}{E}(E_n(\vec{k})) \pd{E_n}{k_l}(\vec{k}) \pd{E_n}{k_m}(\vec{k}) }{\vec{k}}.
\end{equation*}
The remaining terms are then known as the regular, or interband, conductivity
\begin{align} \label{eq:regular}
        &\sigma^{\text{R}}_{lm}(\omega) = - \frac{e^2}{\hbar^2 (2 \pi)^d} \sum_{n \neq n' \in \mathbb{N}_{> 0}} \inty{\Gamma^*}{}{}{\vec{k}}  \\ \nonumber
        &\frac{ \left( \Phi(E_{n'}(\vec{k})) - \Phi(E_n(\vec{k})) \right) \ip{ \chi_{n}(\vec{k}) }{ \left( \pd{H}{k_l}(\vec{k}) \right) \chi_{n'}(\vec{k}) } \ip{ \chi_{n'}(\vec{k}) }{ \left( \pd{H}{k_m}(\vec{k}) \right) \chi_n(\vec{k}) } }{(E_{n'}(\vec{k}) - E_n(\vec{k}))\left(\frac{i}{\hbar}\left(E_{n'}(\vec{k}) - E_n(\vec{k})\right) - i \omega + \Gamma\right)}.
\end{align}

\subsection{Particles in periodic potential in the tight-binding limit} \label{sec:tight-binding}

We now consider the case of periodic tight-binding models. Since tight-binding models approximate continuum models, it should be possible to derive the Kubo formula for tight-binding models directly from \eqref{eq:general_periodic}. The idea would be to approximate the inner products of Bloch functions in \eqref{eq:general_periodic} by linear combinations of inner products of atomic orbitals, following, e.g., \cite{ashcroft_mermin,https://doi.org/10.48550/arxiv.2006.08025,https://doi.org/10.48550/arxiv.2010.12097,https://doi.org/10.48550/arxiv.2107.09146,Fefferman2018,Helffer1984,10.1007/3-540-51783-9_19}. Here we provide a direct derivation, starting from a discrete model.

Let $\Lambda, \Gamma, A, \Lambda^*, \Gamma^*, B$ be as in Section \ref{sec:periodic}. We initially consider the infinite Hilbert space
\begin{equation}
    \mathcal{H}_{\text{inf}} := \ell^2\left( \Lambda ; \mathbb{C}^N \right), \quad \mathcal{H}_{\text{inf}} \ni \psi = \left( \psi_{\vec{R}} \right)_{\vec{R} \in \Lambda}, \quad \psi_{\vec{R}} = \left( \psi_{\vec{R}}^\alpha \right)_{1 \leq \alpha \leq N},
\end{equation}
and local Hamiltonians satisfying
\begin{equation}
    \left( H \psi \right)_{\vec{R}} = \sum_{\vec{R}' \in \Lambda} H_{\vec{R},\vec{R}'} \psi_{\vec{R}'}, \quad \vert H_{\vec{R},\vec{R}'} \vert \leq C e^{- \gamma \vert \vec{R} - \vec{R}' \vert}.
\end{equation}
We assume further that $H$ is periodic, so that
\begin{equation}
    H_{\vec{R} + \vec{v},\vec{R}' + \vec{v}} = H_{\vec{R},\vec{R}'}, \quad \vec{R}, \vec{R}', \vec{v} \in \Lambda.
\end{equation}

Recall $\Lambda_L$ and $\Gamma_L^*$, the truncated lattice and discretized Brillouin zone. We can, then, consider the restriction of $H$ to the truncated Hilbert space
\begin{equation}
    \mathcal{H} := \ell^2\left( \Lambda_L ; \mathbb{C}^N \right), \quad \mathcal{H} \ni \psi = \left( \psi_{\vec{R}} \right)_{\vec{R} \in \Lambda_L}, \quad \psi_{\vec{R}} = \left( \psi_{\vec{R}}^\alpha \right)_{1 \leq \alpha \leq N},
\end{equation}
where we identify points in $\Lambda$ related by vectors in $L \Lambda$. A basis of $\mathcal{H}$ is provided by the discrete Bloch functions
\begin{equation} \label{eq:discrete_Bloch}
    \left\{ \frac{e^{i \vec{k} \cdot \vec{R}} \chi_n(\vec{k})}{L^{\frac{d}{2}}} : \vec{k} \in \Gamma_L^*, 1 \leq n \leq N \right\},
\end{equation}
where the $\chi$s are eigenvectors of the Bloch Hamiltonian 
\begin{equation}
    H(\vec{k}) := \sum_{\vec{R}' \in \Lambda} H_{\vec{0}\vec{R}'} e^{i \vec{k} \cdot \vec{R}'}.
\end{equation}
Evaluating the trace of the Fermi-Dirac distribution with respect to the basis \eqref{eq:discrete_Bloch} to derive $N$ and $\tilde{N}$ proceeds exactly as in Section \ref{sec:periodic}, except that the sum over $n$ is now a finite sum from $1$ to $N$. 

To evaluate the trace formula, we again start by evaluating the matrix elements
\begin{equation}
    \begin{split}
        &\frac{1}{L^d} \sum_{\vec{R} \in \Lambda_L} \overline{ e^{i \vec{k} \cdot \vec{R}} \chi_n(\vec{k}) } \sum_{\vec{R}' \in \Lambda_L} ( \de_l H )_{\vec{R} \vec{R}'} e^{i \vec{k}' \cdot \vec{R}'} \chi_{n'}(\vec{k}')   \\
        &= \frac{1}{L^d} \sum_{\vec{R} \in \Lambda_L} e^{- i \vec{k} \cdot \vec{R}} \overline{ \chi_n(\vec{k}) } \sum_{\vec{R}' \in \Lambda_L} i (R' - R)_l H_{\vec{R} \vec{R}'} e^{i \vec{k}' \cdot \vec{R}'} \chi_{n'}(\vec{k}').
    \end{split}
\end{equation}
Changing variables in the sum over $\vec{R}'$ to $\vec{R}''$, where $\vec{R}' = \vec{R} + \vec{R}''$, we find 
\begin{equation}
    \begin{split}
        &= \frac{1}{L^d} \sum_{\vec{R} \in \Lambda_L} e^{i (\vec{k}' - \vec{k}) \cdot \vec{R}} \overline{ \chi_n(\vec{k}) } \sum_{\vec{R}' \in \Lambda_L} i {R}_l'' H_{\vec{R} \vec{R} + \vec{R}''} e^{i \vec{k}' \cdot \vec{R}''} \chi_{n'}(\vec{k}')    \\
        &= \frac{1}{L^d} \sum_{\vec{R} \in \Lambda_L} e^{i (\vec{k}' - \vec{k}) \cdot \vec{R}} \overline{\chi_n(\vec{k})} \left( \pd{H}{k_l}(\vec{k}) \right) \chi_{n'}(\vec{k}').
    \end{split}
\end{equation}
It is now straightforward to see that
\begin{equation}
    \begin{split}
        &\frac{1}{L^d} \sum_{\vec{R} \in \Lambda_L} \overline{ e^{i \vec{k} \cdot \vec{R}} \chi_n(\vec{k}) } \sum_{\vec{R}' \in \Lambda_L} ( \de_l H )_{\vec{R} \vec{R}'} e^{i \vec{k}' \cdot \vec{R}'} \chi_{n'}(\vec{k}')     \\
        &= \delta_{\vec{k}\vec{k}'} \overline{ \chi_n(\vec{k}) } \left( \pd{H}{k_l}(\vec{k}) \right) \chi_{n'}(\vec{k}).
    \end{split}
\end{equation}
Similar manipulations show that \eqref{eq:general_periodic} also holds for tight-binding models, with the modifications as above.


\subsection{Nearest-neighbor tight-binding model of graphene} \label{sec:graphene}

In this section we consider the nearest-neighbor tight-binding model of graphene; for further discussion of this model, see, e.g., \cite{2009Castro-NetoGuineaPeresNovoselovGeim,Fefferman2018}. We will show that the conductivity of this model can be written as a two-dimensional integral with explicit integrand; see \eqref{eq:graphene_Drude}-\eqref{eq:graphene_regular}. In a previous work \cite{MargetisWatsonLuskin2023}, we used an analogous integral representation for the conductivity to derive asymptotic formulas for the interband conductivity of the one-dimensional SSH model \cite{Su1979}.

The graphene Bravais lattice vectors are 
\begin{equation}
    \vec{a}_1 := \frac{a}{2} \left( 1, \sqrt{3} \right)^\top, \quad \vec{a}_2 := \frac{a}{2} \left( -1 , \sqrt{3} \right)^\top, \quad A := \begin{pmatrix} \vec{a}_1, \vec{a}_2 \end{pmatrix},
\end{equation}
where $a > 0$ is the lattice constant. 
Within the $\vec{R}$th fundamental cell of the lattice, there are two atoms, at positions $\vec{R} + \vec{\tau}^A$ and $\vec{R} + \vec{\tau}^B$, which we will take as 
\begin{equation}
    \vec{\tau}^A := \vec{0}, \quad \vec{\tau}^B := \left(0, d\right)^\top, \quad d := \frac{a}{\sqrt{3}}.
\end{equation}
The reciprocal lattice vectors are
\begin{equation}
    \vec{b}_1 := \frac{4 \pi}{3 d} \left( \frac{\sqrt{3}}{2},\frac{1}{2} \right)^\top, \quad \vec{b}_2 := \frac{4 \pi}{3 d} \left( - \frac{ \sqrt{3} }{2},\frac{1}{2} \right)^\top, \quad B := \begin{pmatrix} \vec{b}_1, \vec{b}_2 \end{pmatrix}.
\end{equation}

We consider one orbital per atom, so that there are $N = 2$ orbitals per unit cell. In the $L \rightarrow \infty$ limit, wave-functions of electrons in graphene are elements of $\ell^2(\Lambda ; \mathbb{C}^2)$, written as $\psi = \left( \psi_{\vec{R}} \right)_{\vec{R} \in \Lambda} = \left( \psi^A_{\vec{R}}, \psi^B_{\vec{R}} \right)^\top_{\vec{R} \in \Lambda}$, where $\vert\psi^\sigma_{\vec{R}}\vert^2$ represents the electron density on sublattice $\sigma \in \{A,B\}$ in the $\vec{R}$th cell. The graphene tight-binding Hamiltonian with nearest-neighbor hopping acts as
\begin{equation} \label{eq:nearest_neighbor}
    \left( H \psi \right)_{\vec{R}} = - t \begin{pmatrix} \psi_{\vec{R}}^B + \psi_{\vec{R}-\vec{a}_1}^B + \psi_{\vec{R}-\vec{a}_2}^B \\ \psi_{\vec{R}}^A + \psi^A_{\vec{R} + \vec{a}_1} + \psi^A_{\vec{R} + \vec{a}_2} \end{pmatrix}, \quad \vec{R} \in \Lambda,
\end{equation}
where $t > 0$ is the nearest-neighbor hopping energy. 

The Bloch Hamiltonian is given explicitly by
\begin{equation} \label{eq:graphene_block}
    H(\vec{k}) := - t \begin{pmatrix} 0 & F(\vec{k}) \\ \overline{F(\vec{k})} & 0 \end{pmatrix}, \quad F(\vec{k}) := 1 + e^{- i \vec{k} \cdot \vec{a}_1} + e^{- i \vec{k} \cdot \vec{a}_2},
\end{equation}
and can be explicitly diagonalized, with eigenpairs 
\begin{equation} \label{eq:monolayer_bands}
    E_\pm(\vec{k}) := \pm t \vert F(\vec{k})\vert, \quad \chi_\pm(\vec{k}) := \frac{1}{\sqrt{2}} \left( 1 , \mp \frac{\overline{F(\vec{k})}}{ \vert F(\vec{k}) \vert } \right)^\top.
\end{equation}

We can now evaluate the diagonal matrix elements
\begin{equation}
    \ip{ \chi_{\pm}(\vec{k}) }{ \left( \pd{H}{k_l}(\vec{k}) \right) \chi_{\pm}(\vec{k}) } = \pm t \frac{ \Re \left( \overline{F(\vec{k})} \pd{F}{k_l}(\vec{k}) \right) }{ \vert F(\vec{k})\vert },
\end{equation}
and off-diagonal matrix elements
\begin{equation}
    \ip{ \chi_{\mp}(\vec{k}) }{ \left( \pd{H}{k_l}(\vec{k}) \right) \chi_{\pm}(\vec{k}) } = \pm t i \frac{ \Im \left( \overline{F(\vec{k})} \pd{F}{k_l}(\vec{k}) \right) }{ \vert F(\vec{k})\vert },
\end{equation}
from which follow formulas for the Drude conductivity
\begin{equation} \label{eq:graphene_Drude}
    \begin{split}
        &\sigma^\text{D}_{lm}(\omega) = - \frac{e^2 t^2}{\hbar^2 (2 \pi)^2 (\Gamma - i \omega)} \sum_{s \in \pm} \inty{\Gamma^*}{}{}{\vec{k}}  \\
        &\fd{\Phi}{E}(s \vert F(\vec{k})\vert) \left( \frac{ \Re \left( \overline{F(\vec{k})} \pd{F}{k_l}(\vec{k}) \right) }{ \vert F(\vec{k})\vert } \right) \left( \frac{ \Re \left( \overline{F(\vec{k})} \pd{F}{k_m}(\vec{k}) \right) }{ \vert F(\vec{k})\vert } \right),
    \end{split}
\end{equation}
and regular conductivity
\begin{equation} \label{eq:graphene_regular}
    \begin{split}
        &\sigma^\text{R}_{lm}(\omega) = - \frac{e^2 t^2}{\hbar^2 (2 \pi)^2} \sum_{s,s' \in \pm, s \neq s'} \inty{\Gamma^*}{}{}{\vec{k}}   \\
        &\frac{ \left( \Phi(s'\vert F(\vec{k})\vert) - \Phi(s\vert F(\vec{k})\vert) \right)}{(s' \vert F(\vec{k}) \vert - s \vert F(\vec{k}) \vert) (s' \vert F(\vec{k}) \vert - s \vert F(\vec{k}) \vert - i \omega + \Gamma)}  \\
        &\times \left( \frac{ \Im \left( \overline{F(\vec{k})} \pd{F}{k_l}(\vec{k}) \right) }{ \vert F(\vec{k})\vert } \right) \left( \frac{ \Im \left( \overline{F(\vec{k})} \pd{F}{k_m}(\vec{k}) \right) }{ \vert F(\vec{k})\vert } \right).
    \end{split}
\end{equation}
A similar calculation can be carried out for the Haldane model~\cite{Haldane1988} (for reviews, see \cite{2013FruchartCarpentier,2018MarcelliMonacoMoscolariPanati,2023ColbrookHorningThickeWatson}). Remarkably, for parameter ranges such that the model has a bandgap, direct calculation from \eqref{eq:regular2} shows that this model exhibits quantized transverse conductivity at zero temperature, dissipation, and frequency (this phenomenon was first observed in the context of the quantum Hall effect \cite{1982ThoulessKohmotoNightingaledenNijs}). This calculation lies beyond our present scope.

\section{Derivation of classical Drude conductivity} \label{sec:classical_formalism}

\subsection{Assumptions and result} \label{sec:classical_assumptions}

In this section, we show how the formalism introduced in Section \ref{sec:results} can be adapted to classical systems to derive the classical Drude conductivity \eqref{eq:free_Kubo}. The main difference between the present derivation and standard derivations (see, for example, Chapter 1 of \cite{ashcroft_mermin}) is that we work with the phase space density, which is the natural analog of the quantum density matrix. Working with the phase space density allows us to provide a derivation which more closely parallels the derivation of the quantum Kubo formula \eqref{eq:standard_Kubo} than standard approaches. We expect that the present classical formalism would emerge naturally from the quantum formalism via a semiclassical limit, with, for example, the time evolution of the density matrix by the von Neumann equation being replaced by the time evolution of the phase space density by the Liouville evolution.

For simplicity, we consider non-interacting, negatively charged, classical particles, with classical Hamiltonian
\begin{equation}
    H_0(\vec{p}) := \frac{p^2}{2 m}.
\end{equation}
We restrict the system to a $d$-dimensional box, with sides of length $L$, and periodic boundary conditions. We work with the phase space density function $\rho(\vec{p},\vec{q},t)$, where $\vec{p}$ and $\vec{q}$ denote $d$-dimensional particle momentum and position, respectively. \\

\noindent \emph{Equilibrium phase space density} At $t = 0$, we assume the system to be at equilibrium, with equilibrium phase space density given by the Maxwellian
\begin{equation} \label{eq:IC}
    \rho(\vec{p},\vec{q},0) = \Phi_M(\vec{p}) := \tilde{N} \left( \frac{ \beta }{ 2 \pi m } \right)^{\frac{d}{2}} e^{- \beta H_0(\vec{p}) }.
\end{equation}
Here, $\beta := (k T)^{-1}$ is the inverse temperature $T$ scaled by Boltzmann's constant $k$, $m$ denotes the particle mass, and $\tilde{N}$ is the particle density. To see that $\tilde{N}$ represents the particle density, we can verify that it equals the total particle number $N$ normalized by the volume $L^d$, since
\begin{equation} \label{eq:N}
    \frac{N}{L^d} = \tilde{N}, \text{ where } N := \inty{\left[-\frac{L}{2},\frac{L}{2}\right]^d}{}{ \inty{\mathbb{R}^d}{}{ \Phi_M(\vec{p}) }{\vec{p}} }{\vec{q}}.
\end{equation}

\begin{remark}
    We take the Maxwellian \eqref{eq:IC} as the equilibrium distribution in this section because our intent is to show that the main ideas of Sections \ref{sec:results} and \ref{sec:quantum_derivation} have close analogs which are entirely classical. If our goal was to give an accurate model of electrons in a real material, it would make sense to replace \eqref{eq:IC} by the Fermi-Dirac distribution \eqref{eq:Fermi-Dirac}. This approach is known as the Sommerfeld theory; see, e.g., \cite{ashcroft_mermin}.
\end{remark}

\noindent \emph{Evolution of phase space density under applied field} For $t > 0$, and in the absence of dissipation (see below), we assume that the particle density evolves according to the von Neumann equation
\begin{equation} 
    \pd{\rho}{t} + \left( \pd{\rho}{\vec{q}} \cdot \pd{H_{\vec{E}}}{\vec{p}} + \pd{\rho}{\vec{p}} \cdot \pd{H_{\vec{E}}}{\vec{q}} \right) = 0, \quad \rho(\vec{p},\vec{q},0) = \Phi_M(\vec{p}),
\end{equation}
where $H_{\vec{E}}$ is the classical Hamiltonian perturbed by an electric field
\begin{equation}
    H_{\vec{E}}(\vec{p},\vec{q},t) := H_0(\vec{p}) + e \vec{E}(t) \cdot \vec{q}.
\end{equation}
Evaluating the partial derivatives of $H_{\vec{E}}$ we obtain
\begin{equation} \label{eq:Liouville}
    \pd{\rho}{t} + \left( \pd{\rho}{\vec{q}} \cdot \frac{\vec{p}}{m} - e \pd{\rho}{\vec{p}} \cdot \vec{E}(t) \right) = 0, \quad \rho(\vec{p},\vec{q},0) = \Phi_M(\vec{p}).
\end{equation}
The solution of the initial value problem \eqref{eq:Liouville} is explicit, given by 
\begin{equation}
    \rho(\vec{p},t) = \Phi_M\left(\vec{p} + e \inty{0}{t}{ \vec{E}(t') }{t'} \right).
\end{equation}
In particular, the particle density $\tilde{N}$ \eqref{eq:N} is conserved by the evolution. \\

\noindent \emph{Modeling of dissipation} We introduce dissipation as follows. We assume that scattering events occur at a random sequence of times $0 = t_0 < t_1 < ...$, such that the differences $\tau_n := t_{n+1} - t_n > 0$ are modeled by a Poisson distribution with rate $\Gamma > 0$, so that
\begin{equation} \label{eq:poisson_classical}
    \mathbb{P}(\tau_n \leq \nu) = \inty{0}{\nu}{ \Gamma e^{- \Gamma \tau_n} }{\tau_n}.
\end{equation}
The average time between scattering events is thus $\frac{1}{\Gamma}$. After each scattering event, we assume that the system returns to equilibrium, before evolving again according to \eqref{eq:Liouville} in between scattering events, i.e., we assume that $\rho$ evolves according to
\begin{equation} \label{eq:Liouville_scattering}
    \pd{\rho}{t} + \left( \pd{\rho}{\vec{q}} \cdot \frac{\vec{p}}{m} - e \pd{\rho}{\vec{p}} \cdot \vec{E}(t) \right) = 0, \quad \rho(\vec{p},\vec{q},t_n) = \Phi_M(\vec{p}),
\end{equation}
for each interval $t_n \leq t < t_{n+1}$, $n \in \mathbb{N}_{\geq 0}$. The solution of \eqref{eq:Liouville_scattering} is again explicit, given by
\begin{equation} \label{eq:explicit_between}
    \rho(\vec{p},t) = \Phi_M\left(\vec{p} + e \inty{t_n}{t}{ \vec{E}(t') }{t'}\right).
\end{equation}

\noindent \emph{Current observable} Our results involve long time averages, averaged over scattering events, of the current density. The current density is defined by
\begin{equation} \label{eq:current}
    \vec{J}(t) := - \frac{e}{m L^d} \inty{\left[-\frac{L}{2},\frac{L}{2}\right]^d}{}{ \inty{\mathbb{R}^d}{}{ \vec{p} \rho(\vec{p},\vec{q},t) }{\vec{p}} }{\vec{q}}.
\end{equation}
It is straightforward to verify that the current density vanishes at equilibrium, i.e., 
\begin{equation}
    - \frac{e}{m L^d} \inty{\left[-\frac{L}{2},\frac{L}{2}\right]^d}{}{ \inty{\mathbb{R}^d}{}{ \vec{p} \Phi_M(\vec{p}) }{\vec{p}} }{\vec{q}} = 0,
\end{equation}
since $\Phi_M(-\vec{p}) = \Phi_M(\vec{p})$. In between scattering events, i.e., within each interval $t_n \leq t < t_{n+1}$, using the explicit formula \eqref{eq:explicit_between}, we have
\begin{equation} \label{eq:unaveraged}
    \vec{J}(t) = \frac{e^2 \tilde{N}}{m} \inty{t_n}{t}{ \vec{E}(t') }{t'}.
\end{equation}

\noindent \emph{Result} For simplicity, we now assume that the applied field is time-harmonic with frequency $\omega \in \mathbb{R}$, so that
\begin{equation} \label{eq:classical_freq}
    \vec{E}(t) = e^{- i \omega t} \vec{E}(\omega). 
\end{equation}
Note that we could allow for a phase $e^{- i \theta}$ in the applied field. Since we do not average over phases in the classical derivation, the result would be to multiply the current by a phase without any other change. In particular, the conductivity would be unaffected. 
We then consider the time-averaged current density at frequency $\omega$, averaged over scattering times distributed according to \eqref{eq:poisson_classical}, as the number of scattering events goes to infinity 
\begin{equation}  \label{eq:limit_classical_0}
    \langle \vec{J} \rangle := \lim_{n \rightarrow \infty} \left\langle \frac{1}{t_n} \inty{0}{t_n}{ e^{i \omega t'} \vec{J}(t') }{t'} \right\rangle_\Gamma,
\end{equation}
where $\langle \cdot \rangle_\Gamma$ denotes the average over $\tau_0, ... , \tau_{n-1}$, each distributed according to \eqref{eq:poisson_classical}. We \emph{assume} that this limit can be calculated, through a Tauberian theorem, by the sequence of limits 
\begin{equation} \label{eq:limit_classical}
    \langle \vec{J} \rangle(\omega) := \lim_{\delta \downarrow 0} \delta \lim_{n \rightarrow \infty} \left\langle \inty{0}{t_n}{ e^{(i \omega - \delta) t'} \vec{J}(t') }{t'} \right\rangle_\Gamma.
\end{equation}
Our result is then the following formula, known as the Drude conductivity
\begin{equation} \label{eq:classical_Drude}
    \langle \vec{J} \rangle(\omega) = \frac{e^2 \tilde{N}}{m (\Gamma - i \omega)} \vec{E}(\omega).
\end{equation}

\subsection{Derivation of classical Drude conductivity} \label{sec:classical_derivation}

We first consider the case $\omega = 0$, writing $\langle \vec{J} \rangle := \langle \vec{J} \rangle(0)$ for simplicity. We start by writing \eqref{eq:limit_classical_0} using a Tauberian theorem as
\begin{equation} \label{eq:averaged_intervals_0}
    \begin{split}
        \langle \vec{J} \rangle &= \lim_{\delta \downarrow 0} \lim_{n \rightarrow \infty} \left\langle \sum_{m = 0}^{n-1} \delta \inty{t_m}{t_{m+1}}{ e^{- \delta t'} \vec{J}(t') }{t'} \right\rangle_{\Gamma}  \\
        &= \frac{e^2 \tilde{N}}{m} \lim_{\delta \downarrow 0} \lim_{n \rightarrow \infty} \left\langle  \sum_{m = 0}^{n-1} \delta \inty{t_m}{t_{m+1}}{ e^{- \delta t'} (t' - t_m) }{t'} \right\rangle_{\Gamma} \vec{E}  \\
        &= \frac{e^2 \tilde{N}}{m} \lim_{\delta \downarrow 0} \lim_{n \rightarrow \infty} \left\langle \sum_{m = 0}^{n-1} \delta e^{- \delta t_m} \inty{0}{\tau_m}{ e^{- \delta t'} t' }{t'} \right\rangle_{\Gamma} \vec{E},
    \end{split}
\end{equation}
using the explicit formula for $\vec{J}$ between scattering events \eqref{eq:unaveraged}. Let us focus on the $m = 0$ term
\begin{equation}
    \delta \inty{0}{\tau_0}{ e^{- \delta t'} t' }{t'} = \frac{1}{\delta} \left( 1 - (1 + \tau_0 \delta) e^{- \delta \tau_0} \right).
\end{equation}
Averaging over $\tau_0$ we find
\begin{equation}\begin{split}
    \frac{1}{\delta} \left( 1 - \inty{0}{\infty}{ \Gamma (1 + \tau_0 \delta)  e^{- (\Gamma + \delta) \tau_0} }{\tau_0} \right) &= \frac{1}{\delta} \left( 1 - \frac{\Gamma}{\Gamma + \delta} - \frac{\Gamma \delta}{(\Gamma + \delta)^2} \right)\\
    &=\frac1{\Gamma+\delta}\left(1-\frac\Gamma{\Gamma+\delta}\right).
\end{split}\end{equation}
The $n = 1$ term is 
\begin{equation}
    \frac{1}{\delta} e^{- \delta \tau_0} \left( 1 - (1 + \tau_1 \delta) e^{- \delta \tau_1} \right).
\end{equation}
Averaging this term over $\tau_0$ and $\tau_1$ we derive 
\begin{equation}
    \frac1{\Gamma+\delta}\left(1-\frac\Gamma{\Gamma+\delta}\right) \left( \frac{\Gamma}{\Gamma + \delta} \right).
\end{equation}
More generally, noting that
\begin{equation} \label{eq:telescoping_0}
    e^{- \delta t_m} = \begin{cases} 1 & m = 0,  \\ \prod_{l = 0}^{m-1} e^{- \delta \tau_l} & m \geq 1, \end{cases}
\end{equation}
we derive
\begin{equation}
    \langle \vec{J} \rangle = \frac{e^2 \tilde{N}}{m} \lim_{\delta \downarrow 0} \frac1{\Gamma+\delta}\left(1-\frac\Gamma{\Gamma+\delta}\right) \lim_{n \rightarrow \infty} \sum_{m = 0}^{n-1} \left( \frac{\Gamma}{\Gamma + \delta} \right)^m.
\end{equation}
Summing the series and taking the limit $n \rightarrow \infty$ we find
\begin{equation}
    \langle \vec{J} \rangle = \frac{e^2 \tilde{N}}{m} \lim_{\delta \downarrow 0} \frac1{\delta}\left(1-\frac\Gamma{\Gamma+\delta}\right).
\end{equation}
Equation \eqref{eq:classical_Drude} with $\omega = 0$ then follows by noting that
\begin{equation}
    \frac{1}{\delta} \left(1-\frac\Gamma{\Gamma+\delta}\right) = \frac{1}{\delta} \left(1-\frac{1}{1+\frac{\delta}{\Gamma}}\right) = \frac{1}{\Gamma} + O(\delta).
\end{equation}

We now consider the case with $\omega \neq 0$. We follow the exact same steps as above
\begin{align} \label{eq:averaged_intervals_1}
        \langle \vec{J} \rangle(\omega) &= \lim_{\delta \downarrow 0} \lim_{n \rightarrow \infty} \left\langle \sum_{m = 0}^{n-1} \delta \inty{t_m}{t_{m+1}}{ e^{(i \omega - \delta) t'} \vec{J}(t') }{t'} \right\rangle_{\Gamma}  \\ \nonumber
        &= \frac{e^2 \tilde{N}}{m (- i \omega)} \lim_{\delta \downarrow 0} \lim_{n \rightarrow \infty} \left\langle \sum_{m = 0}^{n-1} \delta \inty{t_m}{t_{m+1}}{ e^{(i \omega - \delta) t'} \left( e^{- i \omega t'} - e^{- i \omega t_m} \right) }{t'} \right\rangle_{\Gamma} \vec{E}(\omega)  \\
        &= \frac{e^2 \tilde{N}}{m (- i \omega)} \lim_{\delta \downarrow 0} \lim_{n \rightarrow \infty} \left\langle \sum_{m = 0}^{n-1} \delta e^{- \delta t_m} \inty{0}{\tau_m}{ e^{- \delta t'} (1 - e^{i \omega t'}) }{t'} \right\rangle_{\Gamma} \vec{E}(\omega). \nonumber
\end{align}
The $m = 0$ term yields
\begin{equation}
    \begin{split}
        \inty{0}{\tau_0}{ e^{- \delta t} \left( 1 - e^{i \omega t} \right) }{t} &= \left[ \frac{ e^{- \delta t} }{ - \delta } - \frac{ e^{(i \omega - \delta) t} }{ i \omega - \delta } \right]_0^{\tau_0}   \\
        &= \left( \frac{1}{\delta} - \frac{ e^{- \delta \tau_0} }{\delta} \right) - \left( \frac{ e^{ (i \omega - \delta) \tau_0 } }{ i \omega - \delta } - \frac{1}{i \omega - \delta} \right).
    \end{split}
\end{equation}
Averaging over $\tau_0$ we find
\begin{equation}
    \begin{split}
        &\frac{1}{\delta} - \frac{ \Gamma }{\delta (\delta + \Gamma)} - \frac{ \Gamma }{ ( i \omega - \delta )( \delta + \Gamma - i \omega ) } + \frac{1}{i \omega - \delta} \\
        &= \frac{1}{\delta + \Gamma} - \frac{1}{\delta + \Gamma - i \omega}.
    \end{split}
\end{equation}
Again using \eqref{eq:telescoping_0}, the $m$th term in the sum is
\begin{equation}
    \left( \frac{\Gamma}{\Gamma + \delta} \right)^m \left( \frac{1}{\delta + \Gamma} - \frac{1}{\delta + \Gamma - i \omega} \right),
\end{equation}
so we arrive at
\begin{equation}
    \langle \vec{J} \rangle(\omega) = \frac{e^2 \tilde{N}}{m (- i \omega)} \lim_{\delta \downarrow 0} \delta \left( \frac{1}{\delta + \Gamma} - \frac{1}{\delta + \Gamma - i \omega} \right) \lim_{n \rightarrow \infty} \sum_{m = 0}^{n-1} \left( \frac{\Gamma}{\Gamma + \delta} \right)^m.
\end{equation}
Summing the series and taking the limit $n \rightarrow \infty$ we arrive at 
\begin{equation}
    \begin{split}
        \langle \vec{J} \rangle(\omega) &= \frac{e^2 \tilde{N}}{m (- i \omega)} \lim_{\delta \downarrow 0} (\Gamma + \delta) \left( \frac{1}{\delta + \Gamma} - \frac{1}{\delta + \Gamma - i \omega} \right) \\ 
        &= \frac{e^2 \tilde{N} \Gamma}{m (- i \omega)} \left( \frac{1}{\Gamma} - \frac{1}{\Gamma - i \omega} \right),
    \end{split}
\end{equation}
which implies equation \eqref{eq:classical_Drude} with $\omega \neq 0$. 

\section{Conclusion} \label{sec:conclusion}

In this paper, we have provided a self-contained, systematic, formal derivation of a widely applicable form of the Kubo formula for the linear response conductivity, under assumptions sufficient for understanding many electronic properties of materials. We have also shown how many commonly-used forms of the Kubo formula, especially those used in the study of crystalline (periodic atomic structure) materials, arise as special cases of the general formula we derive. We hope that this work will stimulate further research into the mathematical modeling of materials' electronic properties. 

For example, the Kubo formula is useful in the prediction of the asymptotic behavior of the optical (regular) conductivity as a function of frequency in the limit of small energy bandgap, when a symmetry of the electronic Hamiltonian is broken (see, e.g., \cite{MargetisWatsonLuskin2023}). Another interesting direction of research is the accurate numerical calculation of the conductivity in twisted multi-layer graphene, for various  twist angles \cite{Cances2017a,Massatt2020,kubocomp20}. 

\subsection*{Acknowledgments}

The research of the first and third authors (ABW and ML) was supported in part by NSF DMREF Award No. 1922165 and Simons Targeted Grant Award No. 896630.  The second author (DM) is grateful to the School of Mathematics of the University of Minnesota for hosting him as an Ordway Distinguished Visitor in the spring of 2022, when part of this work was completed. The authors wish to thank Tony Low and Tobias Stauber for useful discussions.

\bibliography{library}

\end{document}